\documentclass[english,preprint,showpacs,preprintnumbers,amsmath,amssymb]{revtex4}

\usepackage[T1]{fontenc}
\usepackage[latin9]{inputenc}
\usepackage{color}
\usepackage{units}
\usepackage{amssymb}
\usepackage{graphicx}
\usepackage{esint}
\usepackage{bm}
\usepackage{natbib}
\usepackage{amsmath}
\usepackage{pdfpages}
\usepackage{pgffor}

\usepackage{comment}

\usepackage{ulem}
\setcitestyle{journalcolor= blue}

\usepackage{babel}
\begin{document}
\title{Vanishing thermal equilibration for hole-conjugate fractional quantum Hall states in graphene}

\author{Saurabh Kumar Srivastav}
\affiliation{Department of Physics, Indian Institute of Science, Bangalore 560012, India}
\author{Ravi Kumar}
\affiliation{Department of Physics, Indian Institute of Science, Bangalore 560012, India}
\author{Christian Sp{\r a}nsl{\"a}tt}
\affiliation{Institute for Quantum Materials and Technologies, Karlsruhe Institute of Technology, 76021 Karlsruhe, Germany}
\affiliation{Institut f{\"u}r Theorie der Kondensierten Materie, Karlsruhe Institute of Technology, 76128 Karlsruhe, Germany}
\author{K.Watanabe}
\affiliation{National Institute of Material Science, 1-1 Namiki, Tsukuba 305-0044, Japan}
\author{T. Taniguchi}
\affiliation{National Institute of Material Science, 1-1 Namiki, Tsukuba 305-0044, Japan}
\author{Alexander D. Mirlin}
\affiliation{Institute for Quantum Materials and Technologies, Karlsruhe Institute of Technology, 76021 Karlsruhe, Germany}
\affiliation{Institut f{\"u}r Theorie der Kondensierten Materie, Karlsruhe Institute of Technology, 76128 Karlsruhe, Germany}
\affiliation{Petersburg Nuclear Physics Institute, 188300 St. Petersburg, Russia}
\affiliation{L.\,D.~Landau Institute for Theoretical Physics RAS, 119334 Moscow, Russia}
\author{Yuval Gefen}
\affiliation{Institute for Quantum Materials and Technologies, Karlsruhe Institute of Technology, 76021 Karlsruhe, Germany}
\affiliation{Department of Condensed Matter Physics, Weizmann Institute of Science, Rehovot 76100, Israel}
\author{Anindya Das$^{\dagger}$}
\affiliation{Department of Physics, Indian Institute of Science, Bengaluru 560012, India}


\begin{abstract}
Transport through edge channels is responsible for conduction in quantum Hall (QH) phases. Robust quantized values of charge and thermal conductances dictated by bulk topology appear when equilibration processes become dominant. We report on measurements of electrical and thermal conductances of integer and fractional QH phases, realized in hexagonal boron nitride encapsulated graphite-gated bilayer graphene (BLG) devices for both electron and hole doped sides with different valley and orbital symmetries. Remarkably, for complex edges at filling factors $\nu=\frac{5}{3}$ and $\frac{8}{3}$, closely related to the paradigmatic hole-conjugate  $\nu=\frac{2}{3}$ phase, we find quantized thermal conductance whose values ($3\kappa_{0}T$ respectively $4\kappa_{0}T$, where $\kappa_{0}T$ is the thermal conductance quantum) are markedly inconsistent with the values dictated by topology ($1\kappa_{0}T$ and $2\kappa_{0}T$, respectively). The measured thermal conductance values remain insensitive to different symmetries suggesting its universal nature. Our findings are  supported by a theoretical analysis, which indicates that whereas electrical equilibration at the edge is established over a finite length scale, the thermal equilibration length diverges for strong electrostatic interaction. Our results elucidate the subtle nature of crossover from coherent, mesoscopic to topology-dominated transport.
\end{abstract}

\maketitle 

\begin{figure*}
\centerline{\includegraphics[width=1.0\textwidth]{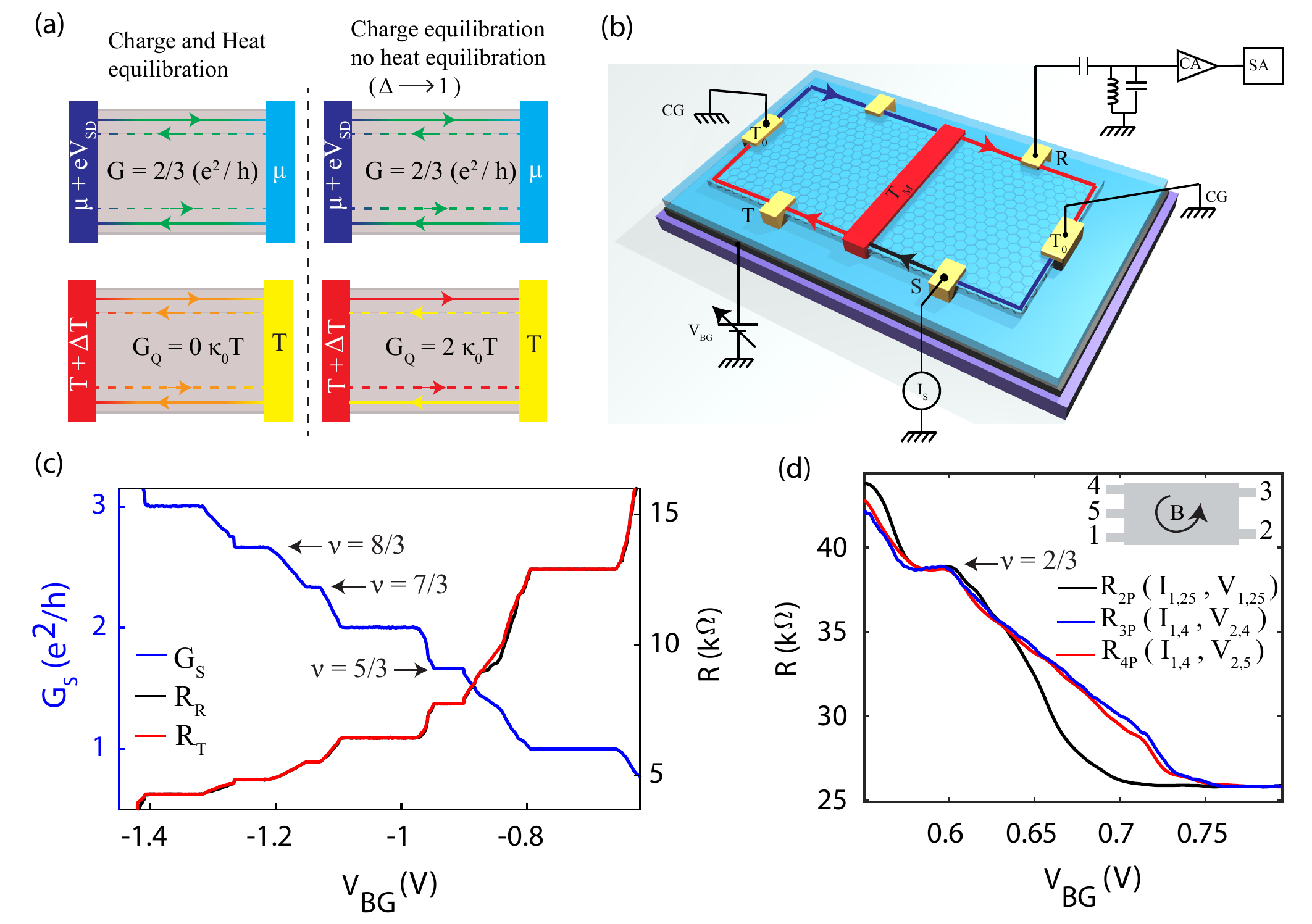}}
\caption{\textbf{Equilibration, Device schematic and QH response.} (\textbf{a}) Left panel: voltage (top) and temperature (bottom) profile in colours with changing intensity along the edge for $L \gg \ell_{\rm eq}^{C}; \ell_{\rm eq}^{H}$. In this limit, $G$ is $\frac{2}{3}\frac{e^2}{h}$ and $G_{Q}$ goes to zero diffusively. Right panel: voltage and temperature profile in limit $\ell_{\rm eq}^{C} \ll L \ll \ell_{\rm eq}^{H}$ realized at $\vartriangle\longrightarrow1$. While $G$ is still $\frac{2}{3}\frac{e^2}{h}$, one has now $G_{Q} = 2\kappa _{0}T$. Solid and dashed lines correspond to the downstream and upstream eigenmodes, respectively. (\textbf{b}) Schematic of device with measurement setup. The device is set in the integer QH regime at $\nu = 1$. An injected current $I_{S}$ (black line) is absorbed in the floating reservoir (red contact) and terminates into two cold grounds. 
The electrical and thermal conductances are measured at low frequency (228 Hz) and high frequency ($\sim$ 758kHz with a LCR resonant circuit), respectively. (\textbf{c}) The blue line is the $G_{S}$ ($I_{S}/V_{S}$) as a function of $V_{BG}$ at B = 10T for the D1 device. The red and black lines are the measured resistances (right y axis) at the $T$ and $R$ contacts, respectively. The robust fractional plateaus at $\frac{5}{3}\frac{e^2}{h}$, $\frac{7}{3}\frac{e^2}{h}$, $\frac{8}{3}\frac{e^2}{h}$ with weaker plateau $\sim$ $\frac{4}{3}\frac{e^2}{h}$ clearly visible. (\textbf{d}) The conductance measured in two probe (black), three probe (blue) and four probe (red) configurations are plotted for the D3 device. The inset show the contact positions. The first number in the subscript of $I$ corresponds to current fed contact and the remaining numbers label grounded contacts. The same notation is used for voltage ($V$) measurements.}
\label{Figure1}
\end{figure*}

According to the bulk-edge correspondence principle~\cite{wen1992theory,swingle2012geometric,dubail2012edge}, certain characteristics of gapless edge modes are constrained by the topological order in the gapped bulk. This turns out to be a subtle issue for hole-conjugate fractional quantum Hall (FQH) phases, whose edges are complex, i.e., hosting counter-propagating modes: $n_{d}$ moving downstream (the direction defined by semiclassical skipping orbits at the edge) and $n_{u}$ moving upstream (opposite to downstream). The quantized two-terminal electrical conductance for these states has been predicted to be $G=\nu \frac{e^{2}}{h}$, while the thermal conductance is $G_{Q}=|n_{d}-n_{u}|\kappa_{0}T$. Here, $\nu$ is the bulk Landau-level filling factor, $\kappa_{0}= \frac{\pi^{2}k^{2}_\mathrm{B}}{3h}$, $k_\mathrm{B}$ is the Boltzmann constant, $h$ is the Planck constant,
and $T$ is the temperature~\cite{kane1996thermal,banerjee2017observed,banerjee2018observation}. Observing the quantization of $G$ and $G_{Q}$ requires full equilibration of the counter-propagating  edge modes~\cite{Protopopov2017,Nosiglia2018}.

The paradigmatic example of a complex edge occurs at $\nu  = \frac{2}{3}$ and consists of counter-propagating $1$ (downstream) and $\frac{1}{3}$ (upstream) modes~\cite{macdonald1990edge}. In the presence of disorder and strong electrostatic inter-mode interaction (parametrized by a single parameter $\Delta\to 1$), these bare modes renormalize to one ballistic downstream charge mode with $G=\frac{2}{3}\frac{e^2}{h}$ and one ballistic upstream neutral mode only at the low-temperature ($T\to0$) and infinite-edge-length ($L\to\infty$) limit~\cite{Kane1994}. However, for finite length and at finite temperature, a robust  $G=\frac{2}{3}\frac{e^2}{h}$ requires full equilibration among the bare modes leading to incoherent transport~\cite{Protopopov2017,Nosiglia2018,footnoteIncoherent}. In the opposite limit of coherent, non-equilibrated edge transport, one finds~\cite{Protopopov2017} $G=\frac{4}{3}\frac{e^2}{h}$. Experimental observation of the crossover from $G=\frac{4}{3}\frac{e^2}{h}$ (entirely non-equilibrated) to $G=\frac{2}{3}\frac{e^2}{h}$ (fully equilibrated) has so sofar been reported only in a GaAs/AlGaAs based device~\cite{Cohen2019}. The corresponding crossover length scale $\ell_{\rm eq}^{C}$ defines the electrical equilibration length. Likewise, the thermal equilibration length $\ell_{\rm eq}^{H}$ defines the crossover from a thermally non-equilibrated edge with thermal conductance $G_{Q} = (n_{d}+n_{u})\kappa_{0}T$ to the topologically constrained and equilibrated thermal conductance $G_{Q} = |n_{d}-n_{u}|\kappa_{0}T$. For $\nu  = \frac{2}{3}$ (where $n_{d}=n_{u}=1$) and its cousin states, $\nu  = \frac{5}{3}$ and $\frac{8}{3}$,  this topology-dictated $G_{Q} / \kappa_{0}T$ is 0, 1, and 2, respectively. For $\nu = \frac{2}{3}$, the zero value is expected with increasing $L$ as $G_Q \sim \ell_{\rm eq}^{H}/L$, signalling heat diffusion~\cite{Protopopov2017,Nosiglia2018}. To date, experimental studies of thermal transport on complex FQH edges have been performed only in GaAs/AlGaAs based structures~\cite{banerjee2017observed,banerjee2018observation}, including $\nu=\frac{2}{3}$, and yielded values of $G_{Q}$ consistent with the equilibrated regime. However, similar to the electrical conductance, an experimental manifestation of the dichotomy between equilibrated and non-equilibrated values of heat conductance on complex FQH edges is missing. In this context, a different system like bilayer graphene (BLG) with more degrees of freedom (valley and orbital) together with unprecedented universal edge profile~\cite{hu2011realizing,li2013evolution} due to atomically sharp confining potential are ideal platforms to study the thermal transport. For electron-like FQH edges in this system (with only downstream modes), topology dictated and universal thermal conductance values were found~\cite{Srivastaveaaw5798}, but no measurements have so far been performed for complex hole-conjugate FQH edges.

Here, we report measurements of the thermal and electric conductance of a variety of QH phases, realized in hBN encapsulated graphite gated BLG devices, for both electron and hole doping, using sensitive noise thermometry~\cite{jezouin2013quantum,banerjee2017observed,banerjee2018observation,Srivastaveaaw5798}, where all the symmetries (spin, valley and orbitals) are broken~\cite{li2017even,PhysRevB.98.155421,supplement}. For integer QH ($\nu=1, 2, 3, 4$) and  electron-like FQH states ($\nu= \frac{4}{3}, \frac{7}{3}$) we obtain  the expected values for $G$ ($1\frac{e^2}{h}, 2\frac{e^2}{h}, 3\frac{e^2}{h}, 4\frac{e^2}{h}, \frac{4}{3}\frac{e^2}{h}, \frac{7}{3}\frac{e^2}{h}$, respectively) and $G_Q$ (within accuracy of $5\%$, $1\kappa_{0}T, 2\kappa_{0}T, 3\kappa_{0}T$, $4\kappa_{0}T, 2\kappa_{0}T, 3\kappa_{0}T$, respectively). For the hole-conjugate phases $\nu=\frac{5}{3}$ and $\frac{8}{3}$, $G$ shows expected values ($\frac{5}{3}\frac{e^2}{h}$ and $\frac{8}{3}\frac{e^2}{h}$, respectively), corresponding to electrically equilibrated edges. At the same time, and most remarkably, $G_Q$ is found to be $3\kappa_{0}T$ and $4\kappa_{0}T$, respectively, corresponding to thermally non-equilibrated edges. Our results of thermal conductance on FQH states for different valleys and orbitals further suggest a universality; a topology dictated $G_Q$ for electron-like FQH states ($\frac{4}{3}, \frac{7}{3}$) but entirely non-equilibrated $G_Q$ for hole-conjugate FQH states ($\frac{5}{3}, \frac{8}{3}$). To explain the striking contrast between electric and thermal equilibration for hole-conjugate FQH states, we present a theoretical analysis of edge equilibration in the strong interaction limit  \cite{energyrelax}. In the limit of $\Delta\to1$ we find that, while $\ell_{\rm eq}^{C}$ remains finite, $\ell_{\rm eq}^{H}$ diverges as $1/(\Delta-1)$, indicating vanishing thermal equilibration. This gives rise to a new regime $\ell_{\rm eq}^{C} \ll L \ll\ell_{\rm eq}^{H}$ observed here; in Fig. 1a we contrast it to the regime of fully equilibrated transport, $\ell_{\rm eq}^{C}; \ell_{\rm eq}^{H}\ll L$.

\begin{figure*}
\centerline{\includegraphics[width=1\textwidth]{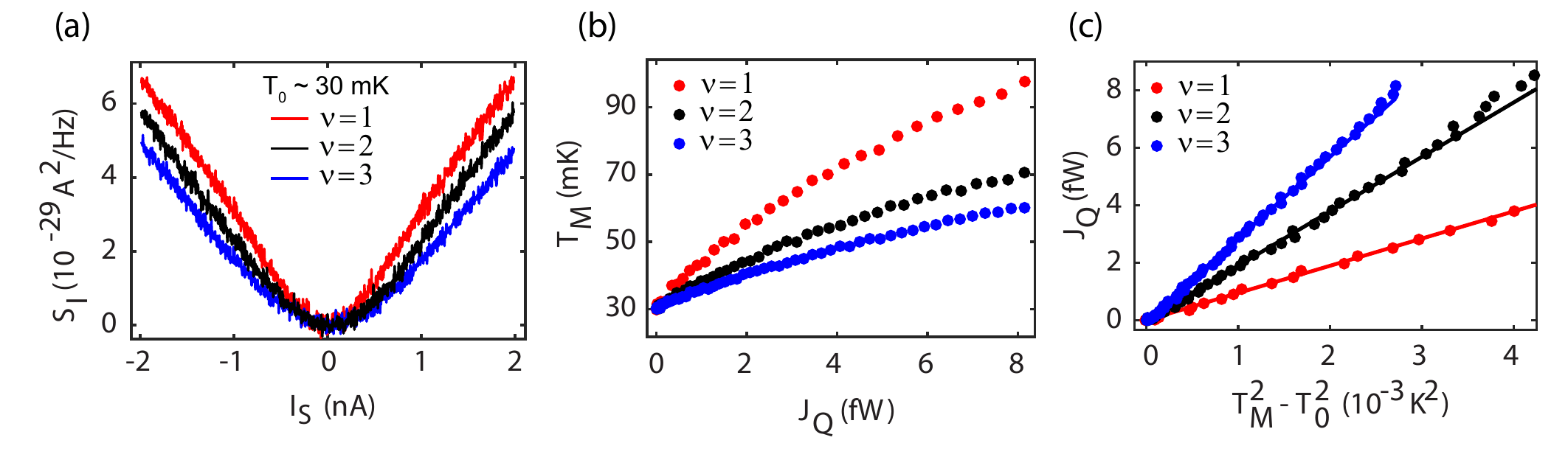}}
\caption{\textbf{Thermal conductance for integer QH states.} (\textbf{a}) Excess thermal noise $S_{I}$ as a function of source current $I_{S}$ at $\nu=1$ (red), 2 (black) and 3 (blue). (\textbf{b}) The temperature $T_{M}$ of the floating contact as a function of the dissipated power $J_{Q}$ for $\nu$ = 1 (red), 2 (black) and 3 (blue), respectively. (\textbf{c}) $J_Q$ (solid circles) as a function of $T^2_{M} - T^2_{0}$ for $\nu$ = 1 (red), 2 (black) and 3 (blue), respectively. Solid lines are linear fits with $G_{Q} = 0.99$, $1.96$ and $3.01\kappa _{0}T$ for $\nu$ = 1, 2 and 3, respectively.}
\label{Figure2}
\end{figure*}

For the thermal conductance measurement, we used two bottom graphite gated devices (D1 and D2), where the graphene was encapsulated between two hBN layers, each with thickness of $\sim$ 20 nm. The device fabrication is described in the Supplemental Material (SM) (\cite{supplement}). Similar to our previous work~\cite{Srivastaveaaw5798}, our devices consist of a floating metallic reservoir in the middle, connected to edge channels on both sides, as shown schematically in Fig. 1(b)~\cite{Srivastaveaaw5798}. The distances from the floating contact to the transverse contacts and cold grounds in Fig. 1(b) were $\sim$ $3\mu m$ ($4\mu m$) and $\sim$ $6\mu m$ ($8\mu m$) for D1 (D2), respectively (see \cite{supplement} for optical images). The electrical conductance was measured using standard lock-in technique whereas the thermal conductance was measured with noise thermometry~\cite{jezouin2013quantum,banerjee2017observed,banerjee2018observation,Srivastaveaaw5798} (\cite{supplement}). In Fig. 1(c), the blue curve represents $G_{S}$ ($I_{S}/V_{S}$) measured at the source contact ($S$) for the D1 device as a function of the bottom graphite gate voltage ($V_{BG}$).
 
Plateaus appear at $\nu =$ $\frac{5}{3}$, $\frac{7}{3}$ and $\frac{8}{3}$ along with the integer QH plateaus at $\nu =1,2,$ and $3$. Similarly, for the D2 device, plateaus appear at $\nu =$ $\frac{4}{3}$, $\frac{7}{3}$ and $\frac{8}{3}$ (\cite{supplement} Fig. S6). In Fig. 1c, the red (black) curve shows the measured resistance $R_{T} = V_{T}/I_{S}$ ($R_{R} = V_{R}/I_{S}$) at the $T$ ($R$) contact along the transmitted (reflected) path from the floating contact (D1). Measured resistances along these paths are exactly half of the resistance measured at the $S$ contact, which strongly suggest that the injected current is equally divided from the floating contact to both sides of the graphene channel. The resistance values at the $S$, $R$ and $T$ contacts for $\nu =$ $\frac{5}{3}$ and $\frac{8}{3}$ are consistent with the charge equilibration of the bare modes along the propagation length. To further confirm the charge equilibration, we measure the conventional two-probe electrical conductance of $\nu = \frac{2}{3}$ in another device (D3 with $L \sim 5-6 \mu m$) with three-probe and four-probe configurations. In Fig. 1d, the quantized value is fixed at $\sim$ 39$k\Omega$. By contrast, if there was no charge equilibration, the resistance values obtained using the Landauer B{\"u}ttiker model~\cite{buttiker1988absence} for our devices will be quite different~\cite{supplement}.

\begin{figure*}
\centerline{\includegraphics[width=1.0\textwidth]{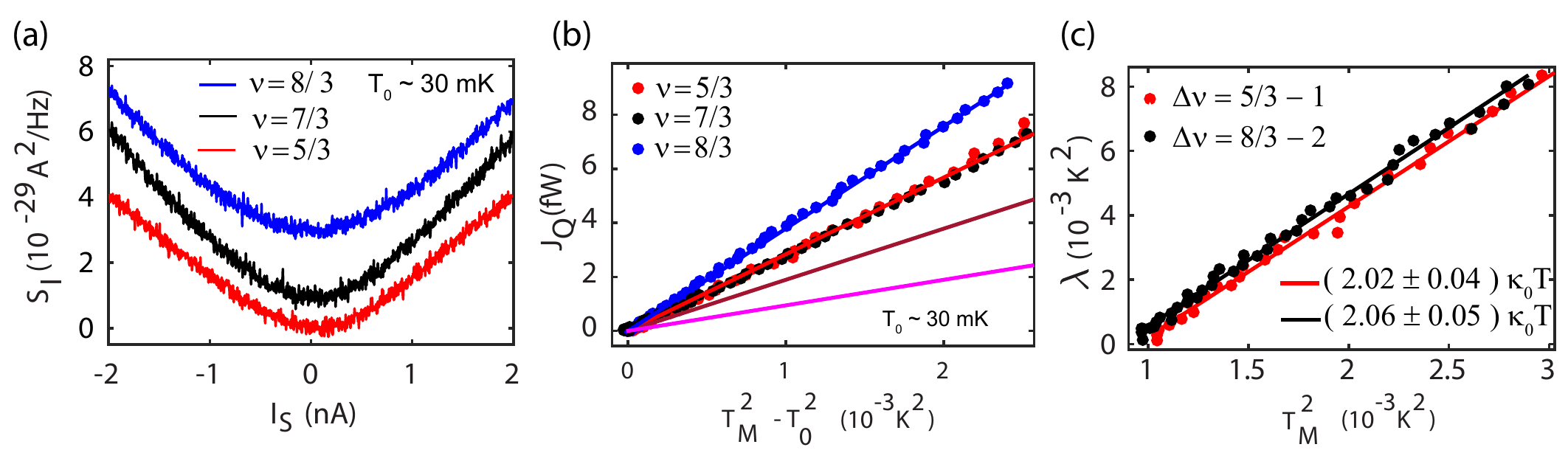}}
\caption{\textbf{Thermal conductance for fractional QH states.} (\textbf{a})  $S_I$ as a function of $I_{S}$ at $\nu$ = 5/3 (red), 7/3 (black) and 8/3 (blue). Experimental curves for 7/3 and 8/3 are shifted vertically by $1\times10^{-29}A^2/Hz$ and $3\times10^{-29}A^2/Hz$. (\textbf{b}) $J_Q$ (solid circles) as a function of $T^2_{M} - T^2_{0}$ for $\nu$ = 5/3 (red), 7/3 (black) and 8/3 (blue). The solid magenta, brown, red and blue lines represent $G_{Q}$ = 1$\kappa _{0}T$, 2$\kappa _{0}T$, 3$\kappa _{0}T$ and 4$\kappa _{0}T$, respectively. The linear fits of the solid circles give $G_{Q}$ = 3.03, 2.96 and 4.03$\kappa _{0}T$ for $\nu$ = 5/3, 7/3 and 8/3, respectively. (\textbf{c}) $\lambda = \Delta J_{Q}/(0.5 \kappa_{0})$ as a function of $T^2_{M}$ for $\Delta \nu = 5/3-1$ (red) and $\Delta \nu = 8/3-2$ (black), where $\Delta J_{Q} = J_{Q}(\nu_{i},T_{M})-J_{Q}(\nu_{j},T_{M})$. Solid lines represent linear fits.  Extracted values of $G_{Q}$ of the $2/3$-like FQH states are $2.02\kappa _{0}T$ and $2.06\kappa _{0}T$ for $\Delta \nu = 5/3-1$ and $\Delta \nu = 8/3-2$, respectively.} 
\label{Figure3}
\end{figure*}

In order to measure the thermal conductance, a DC current ($I_{S}$), injected at the $S$ contact (Fig. 1b), flows towards the floating reservoir and the outgoing current splits into two equal parts to the cold grounds. The power dissipation at the floating reservoir due to joule heating is $J_{Q} = \frac{I_S^2}{4 \nu G_0}$ (SM \cite{supplement}), \cite{Srivastaveaaw5798}, and thus the electrons in the floating reservoir will be heated to a new steady state temperature ($T_{M}$), determined by the following heat balance relation~\cite{jezouin2013quantum,banerjee2017observed,banerjee2018observation,Srivastaveaaw5798,sivan1986multichannel,butcher1990thermal}
\begin{equation}
J_{Q}= J^{e}_{Q}(T_{M},T_{0}) + J^{e-ph}_{Q}(T_{M},T_{0}) \label{eq.heatbalance1}
\end{equation}
\begin{equation}
J_{Q}= 0.5 N \kappa _{0}(T^{2}_{M}-T^{2}_{0}) + J^{e-ph}_{Q}(T_{M},T_{0}) \label{eq.heatbalance2}
\end{equation}
Here, $J^{e}_{Q}(T_{M},T_{0})$ is the electronic contribution of the heat current via $N$ chiral edge modes, and $J^{e-ph}_{Q}(T_{M},T_{0})$ is the heat loss via electron-phonon coupling. The $T_{M}$ is obtained by measuring the excess thermal noise; $S_{I} = \nu k_{B}(T_{M}-T_{0})G_{0}$~\cite{jezouin2013quantum,banerjee2017observed,banerjee2018observation,Srivastaveaaw5798,sivan1986multichannel,jiang2016linear,beenakker1992suppression,blanter2000semiclassical}, along the outgoing edge channels as shown in Fig. 1b. Fig. 2a shows the measured excess thermal noise $S_{I}$ as a function of current $I_{S}$ for $\nu =$ 1 (red), 2 (black), and 3 (blue) (D1). The noise and current axes of Fig. 2a are converted to $J_{Q}$ and $T_{M}$, and plotted in Fig. 2b. To extract $G_{Q}$ for each filling factor, we have plotted $J_{Q}$ as a function of $T^2_{M} - T^2_{0}$ in Fig. 2c. The solid circles represent the experimental data, while the solid lines are the linear fits of $G_{Q}$ with $0.99$, $1.96$, and $3.01\kappa _{0}T$ for $\nu=$1, 2 and 3, respectively. Similarly, for device D2, $G_{Q}$ was found to be $\sim$ $0.99$, $2.05$, $3.04$ and $3.96\kappa _{0}T$ for $\nu=$1, 2, 3 and 4, respectively (\cite{supplement} Fig. S7), which shows an excellent match with its expected theoretical values. Note that $J^{e-ph}$ was negligible up to $T_{M} \sim 60$mK and also, heat Coulomb blockade~\cite{sivre2018heat} was expected to be absent for our graphite gated devices~\cite{supplement}.\\
 
Fig. 3a shows $S_{I}$ as a function of $I_{S}$ for $\nu = \frac{5}{3}$ (red), $\frac{7}{3}$ (black) and $\frac{8}{3}$ (blue) for D1. Experimental curves for 7/3 and 8/3 are shifted vertically by $1\times10^{-29}A^2/Hz$ and $3\times10^{-29}A^2/Hz$ for clarity.  From these raw data, the $T_{M}$ was extracted as a function of $J_{Q}$~\cite{supplement}. In Fig. 3b, $J_{Q}$ is plotted as a function of $T^2_{M} - T^2_{0}$ as shown by the coloured circles, and the solid lines are the theoretical curves for $G_{Q} = 1\kappa _{0}T$ (magenta), $2\kappa _{0}T$ (brown), $3\kappa _{0}T$ (red) and $4\kappa _{0}T$ (blue). The linear fittings to the measured data in Fig. 3b gives $G_{Q} \sim 3.03$, $2.96$, and $4.03\kappa _{0}T$ for $\frac{5}{3}$, $\frac{7}{3}$ and $\frac{8}{3}$, respectively. Similarly, $G_{Q} \sim 1.96$, $3.01$, and $3.94\kappa _{0}T$ for $\frac{4}{3}$, $\frac{7}{3}$ and $\frac{8}{3}$, respectively, for device D2 (\cite{supplement} Fig. S8). 
For the particle-like states $\frac{4}{3}$ and $\frac{7}{3}$, the measured value of $G_{Q}$ is in excellent agreement with the expected theoretical values. However, for the hole-like FQH states $\frac{5}{3}$ and $\frac{8}{3}$, the measured $G_{Q}$ strikingly matches with $(n_{d}+n_{u})\kappa _{0}T$ rather than the expected topological quantum number of $|n_{d}-n_{u}|\kappa _{0}T = 1\kappa _{0}T$, and $2\kappa _{0}T$, respectively. In Fig. 3c, we plot $\lambda = \Delta J_{Q}/(0.5 \kappa_{0})$ as a function of $T^2_{M}$ for two different configurations of $\Delta \nu = \frac{5}{3}-1$ (red) and $\frac{8}{3}-2$ (black) to extract the contribution of the partially filled Landau level with $\nu = $ $\frac{2}{3}$ out of the data for $\frac{5}{3}$ and $\frac{8}{3}$. Linear fits give $2.02\kappa _{0}T$ and $2.06\kappa _{0}T$, respectively, for $G_{Q}$ of the $\nu = $ $\frac{2}{3}$ state. For the D2 device, the fit yields $1.99\kappa _{0}T$ (\cite{supplement} Fig. S8). It is worth to mention here that the thermal conductance of $\frac{7}{3}$ and $\frac{8}{3}$ states observed in hole doped (device D1) and in electron doped (device D2) regime are the same, irrespective of the different orbital nature of their wave-functions, which is N=1(0) in hole(electron) doped regime\cite{li2017even,PhysRevB.98.155421,supplement}. In fact, the extracted thermal conductance of $\frac{2}{3}$ like state from $\frac{5}{3}$ and $\frac{8}{3}$ (Fig. 3(c)) is also the same, irrespective of the different orbital nature of wave-functions, which is N=1(0) for $\frac{5}{3}$($\frac{8}{3}$). This establishes the universality of our results.

The observed values of the thermal conductance for $\nu=\frac{5}{3}$ and $\frac{8}{3}$ imply essentially vanishing thermal equilibration between counter-propagating modes. To explain this, we consider a model of counter-propagating $1$ and $\frac{1}{3}$ modes in the uppermost Landau level. In the presence of inter-channel interactions, this level consists of two emergent, counter-propagating eigenmodes. Their dimensionless charge conductances are $g_{\pm}=(\Delta\pm 1)/3$. Importantly, their dimensionless heat conductances are unity, 
independent of $\Delta$. Tunneling facilitated by random disorder leads to equilibration between these modes. Calculating  the charge and heat tunneling currents, we derive thermal ($\ell^{H}_{\rm eq}$) and charge ($\ell^{C}_{\rm eq}$) equilibration lengths~\cite{supplement};
\begin{equation}
	\ell^{H/C}_{\rm eq} \propto \mathcal{C}^{H/C}(\Delta) T^{2-2\Delta}, \quad \mathcal{C}^H(\Delta) \sim \frac{1}{\Delta-1}, \quad
 \mathcal{C}^C(\Delta)\sim 1,
\end{equation} 
where we have displayed only the dependence on the temperature $T$ and $\Delta$. Our key observation is that the coefficient $\mathcal{C}^H(\Delta)$ diverges for $\Delta \rightarrow 1$, implying a very large $\ell^{H}_{\rm eq}$. This happens because the tunneling current between eigenmodes is proportional to $\Delta-1$. The region $\Delta$ close to $1$ corresponds to very strong interactions. We argue that the sharp 
confining potential of our graphene devices, where the screening graphite gate is separated from the electron gas of graphene by a thin insulating $hBN$ layer ($\sim 10-20$ nm)~\cite{hu2011realizing,li2013evolution,deprez2020tunable,ronen2020aharonov}, favors this regime in contrast to the shallow confining potential in GaAs/AlGaAs devices~\cite{banerjee2017observed,banerjee2018observation}. For $\ell^{C}_{\rm eq}$, the smallness of the tunneling current is compensated by the smallness of the charge conductance of one of the eigenmodes ($g_-$). The eigenmode conductances determine the effect of tunneled charge on the local voltages. Tunneling of a finite charge to the "almost neutral" chiral mode results in an enhanced effect on the local voltage of the mode, facilitating easier equilibration of the local chemical potentials. Technically, this will compensate the $\Delta-1$ factor of the tunneling current, leading to $\mathcal{C}^C(\Delta)\sim 1$ (the same result for $\ell^{C}_{\rm eq}$ is also obtained~\cite{Spanslatt2020Contacts} by explicitly considering electrostatics of fractionalization-renormalized tunneling.)
As a result, for $\Delta $ close to $1$, $\ell^{C}_{\rm eq}\ll \ell^{H}_{\rm eq}$, which creates a broad regime of system sizes $\ell^{C}_{\rm eq}\ll L\ll\ell^{H}_{\rm eq}$ thereby explaining the experimental observations of efficient charge equilibration but vanishing thermal equilibration.\\

In conclusion, the findings of this work are a remarkable manifestation of a transport regime with partial equilibration: the charge transport is in an equilibrated regime, while the heat transport is non-equilibrated irrespective of the  different symmetry nature of wave-functions. Both quantities, in the asymptotic limits of an equilibrated/non-equilibrated edge, respectively, are determined by the edge quantum numbers. We expect that such regimes should be relevant also to other FQH states and materials. In fact, several proposed mechanisms for explaining the observed heat conductance $\frac{5}{2}\kappa_0 T$ at $\nu=5/2$ involve patterns of partial equilibration within the (non-abelian) anti-Pfaffian state~\cite{PhysRevB.99.085309,Simon2020,Asasi2020,park2020noise}. We envisage future work exploring the influence of partial equilibration on noise, decoherence, and FQH interferometry.\\

Useful discussions with I.V. Gornyi, D.G. Polyakov, and I.V. Protopopov on the theory part of this work are acknowledged. 
A.D. thanks the Department of Science and Technology (DST), India for financial support (DSTO-2051) and acknowledges the Swarnajayanti Fellowship of the DST/SJF/PSA-03/2018-19. A.D. also thanks Moty Heiblum and Jainendra Jain for useful discussions. S.K.S. acknowledges PMRF, MHRD for financial support. R. K. acknowledges INSPIRE, DST for financial support. C.S., A.D.M. and Y.G. acknowledge support by DFG Grant No. MI 658/10-1 and by the German-Israeli Foundation Grant No. I-1505-303.10/2019. A.D. and Y.G. acknowledge ICTS, Novel phases of the quantum matter conference for initiation of collaboration. Y.G. acknowledges support by the Helmholtz International Fellow Award. K.W. and T.T. acknowledge support from the Elemental Strategy Initiative conducted by the MEXT, Japan and the CREST (JPMJCR15F3), JST.

S.K.S and R.K. contributed equally to this work. 
\\ $^{\dagger}$Corresponding author.
anindya@iisc.ac.in


\begin{thebibliography}{50}
\bibitem[{foo()}]{footnoteIncoherent}
\bibinfo{journal}{We assume generally that the coherence length is the shortest
  physical length scale governing edge transport. Then, equilibrated transport
  is always incoherent}.
\bibitem[{ene()}]{energyrelax}
\bibinfo{journal}{Our regime is very different from that in
  Ref.~\cite{PhysRevLett.105.056803,PhysRevLett.105.226804,PhysRevLett.125.256803}
  where energy relaxation was much faster than charge relaxation}.
\end{thebibliography}

\newpage
\thispagestyle{empty}
\mbox{}


\section*{Supplementary material}

**SI section-1: Device characterization and noise measurement setup:**

In our experiment, encapsulated devices (heterostructure of hBN/bilayer graphene(BLG)/hBN/graphite) were made using standard dry transfer pick-up technique [1]. Fabrication of these heterostructure involved mechanical exfoliation of hBN and graphite crystals on oxidized silicon wafer using the widely used scotch tape technique. First, a hBN of thickness of $\sim$ 20 nm was picked up at 90°C using a Poly-Bisphenol-A-Carbonate (PC) coated Polydimethylsiloxane (PDMS) stamp placed on a glass slide, attached to tip of a home build micromanipulator. Now this hBN flake was aligned on top of previously exfoliated BLG. BLG was picked up at 90°C. Next step involved the pick up of bottom hBN ($\sim$ 20 nm). This bottom hBN was picked up using the previously picked-up hBN/BLG following the previous process. This hBN/BLG/hBN heterostructure was used to pick-up the graphite flake following previous step. Finally, this resulting heterostructure (hBN/BLG/hBN/graphite) was dropped down on top of an oxidized silicon wafer of thickness 285 nm at temperature 180°C. To remove the residues of PC, these final stacks were cleaned in chloroform ($CHCl_3$) overnight followed by cleaning in acetone and and iso-propyl alcohol (IPA). After this, Poly-methyl-methacrylate (PMMA) photoresist was coated on this heterostructure to define the contacts region in Hall probe geometry using electron beam lithography (EBL). Apart from the conventional Hall probe geometry, we define a region of $\sim$ 6-8 $\mu m^2$ area in the middle of BLG flake, which will act as floating metallic reservoir on edge contact metallization. After EBL, reactive ion etching (mixture of $CHF_3$ and $O_2$ gas with flow rate of 40 sccm and 4 sccm, respectively at 25°C with RF power of 60W) was used to define the edge contact. The etching time was optimized such that the bottom hBN does not etch completely to isolates the contacts from bottom graphite flake, which was used as the back gate. Finally, thermal deposition of Cr/Pd/Au (3/12/60 nm) was done in a evaporator chamber having base pressure of $\sim 1-2\times 10^{-7}$ mbar. After deposition, lift-off procedure was performed in hot acetone and IPA. This results into a Hall bar device along with the floating metallic reservoir connected to the both sides of BLG by the edge contacts. The schematic of the device and measurement set-up are shown in Fig. 1(b). The distances from the floating contact to the transverse contacts and ground contacts were $\sim 3\mu m$ ($4\mu m$) and $\sim 6\mu m$ ($8\mu m$) for D1 (D2), respectively (see Fig. S1 for optical images). All the measurements are done in a cryo-free dilution refrigerator having base temperature of $\sim$ 12mK. Though the electron temperatures ($T_0$) were $\sim$ 30mK and $\sim$ 40mK for D1 and D2, respectively (see SI section S3). Note that in order to see the robust fractional states below $\nu=1$, the D3 device was screened by the both top and bottom graphite gates, and it exhibits clear QH plateaus for electron as well as hole-like FQH states with vanishing longitudinal resistance $R_{xx}$ as shown in Fig. S6. The electrical conductance was measured using standard Lock-in technique where as the thermal conductance was measured employing noise thermometry based on LCR resonant circuit at resonance frequency of $\sim$ 760kHz and amplified by home made preamplifier at 4K followed by room temperature amplifier, and finally measured by a spectrum analyzer.

Two terminal total resistances (R) of D1, D2 and D3 devices were measured as a function of bottom

2

graphite gate voltage ($V_{BG}$) at zero magnetic field. The measured data is fitted with equation[2,3]

$$R = R_C + \frac{L}{We\mu\sqrt{(n_0^2 + (\frac{C_{BG}(V_{BG}-V_{CN})}{e})^2)}} \quad (S1)$$

where $R_C$, L, W, $\mu$, and e are the contact resistance, length, width, mobility, and electron charge, respectively. Carrier concentration of channel is given by $\frac{C_{BG}(V_{BG}-V_{CN})}{e}$ with $C_{BG}$ and $V_{CN}$ are the capacitance per unit area of bottom graphite gate ($\sim$ 20nm hBN) and voltage at the charge neutrality point, respectively. $n_0$ is the charge inhomogeneity.

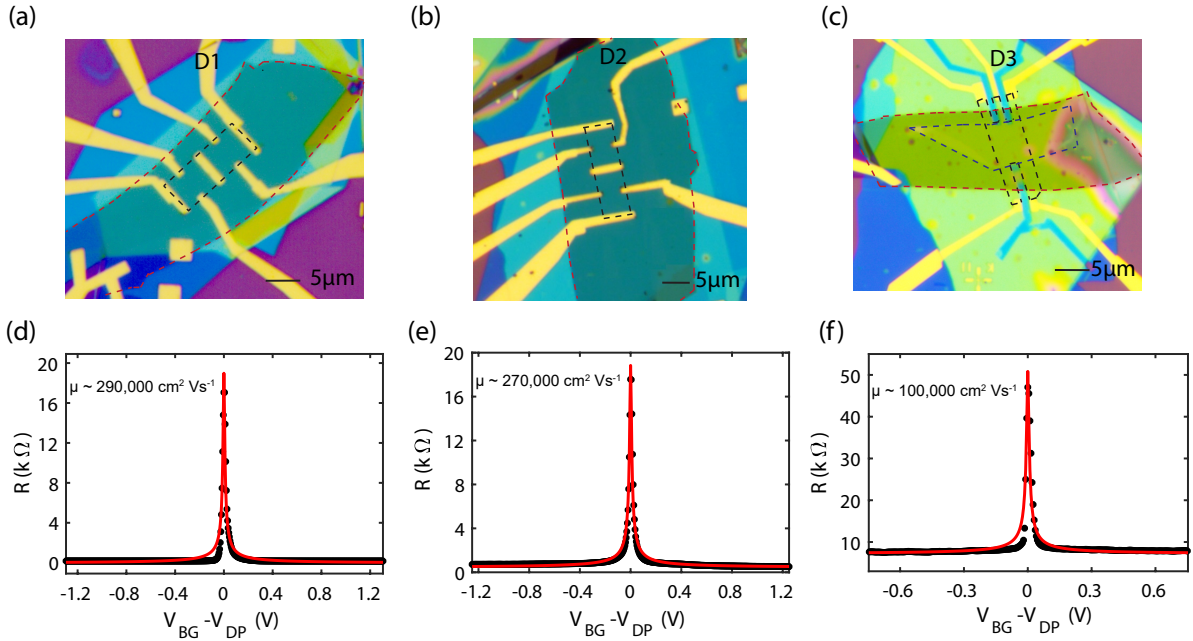

**Fig. S 1: Optical image and device characterization.** Optical images of D1, D2, and D3 devices are shown in (**a**), (**b**) and (**c**), respectively. Region of graphene, bottom graphite, and top graphite are marked by the black, red and blue dashed line respectively. (**d,e,f**) Two probe gate response is plotted as a function of bottom graphite gate voltage at 240 mK. Solid circle shows the experimental data and the red curve is the fit of data in accordance to Eq. (S1). This fit gives the mobility of $\sim$ 290,000 $cm^2V^{-1}s^{-1}$, $\sim$ 270,000 $cm^2V^{-1}s^{-1}$ and 100,000 $cm^2V^{-1}s^{-1}$ of D1, D2 and D3 devices, respectively. The high mobility of the device is necessary to observe fractional states. In these devices, carrier inhomogeneity was found to be order of $5 \times 10^9 cm^{-2}$, which is one order of magnitude smaller than $SiO_2$ gated devices. It should be worth to mention here that although mobility of D3 (top and bottom graphite gated) is smaller than other two, carrier inhomogeneity is in D3 was smaller compared to other two devices, and probably this would be the reason behind the observation of fractional states below integer filling 1.

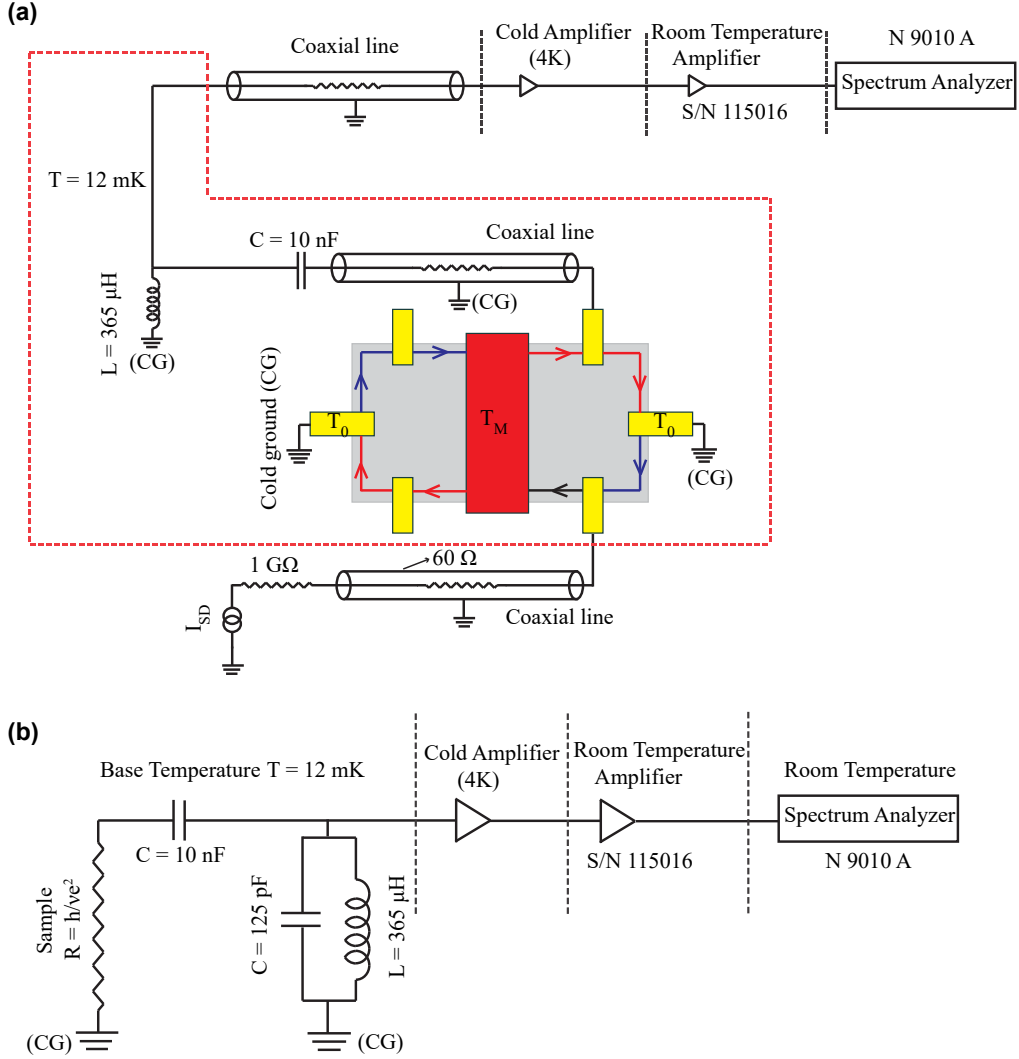

**Fig. S 2: Experimental set-up for noise measurement.** (**a**) Schematic of the measurement set-up. The device was mounted on a chip carrier which was connect to the cold finger fixed to the mixing chamber plate of dilution refrigerator. The ground contact pins are directly shorted to the cold finger to achieve the cold ground. The sample was current biased with a 1 GΩ resistor located at the top of dilution fridge. Current fluctuations measured at contacts located along the reflected current path are converted on chip into voltage fluctuations using the well defined quantum Hall (QH) resistance $R = h/\nu e^2$ (as shown schematically in fig(**b**)) where $\nu$ is the filling factor. The noise signal was amplified with a home made cryogenic voltage pre-amplifier, which was thermalized to 4K plate of dilution refrigerator. This pre-amplified signal was then amplified using a voltage amplifier (PR-E3-SMA S/N 115016) placed at the top of the fridge at room temperature. After second stage of amplification, amplified signal was measured using a spectrum analyzer (N9010A). All the noise measurements were done using the band width of $\sim$ 30 kHz. The resonant L//C tank circuit was built using inductor L of $\sim$ 365 $\mu$H made from a superconducting coil thermally anchored to the mixing chamber of dilution refrigerator. The parallel C of $\sim$ 125 pF is the capacitance that develops along the coaxial lines connecting the sample to the cryogenic pre-amplifier. A ceramic capacitance of 10 nF was introduced between sample and inductor to block the DC current along the measurement line. The typical input voltage noise and current noise of cryogenic pre-amplifiers were $\sim$ 250 $pV/\sqrt{Hz}$ and $\sim$ 20-25 $fA/\sqrt{Hz}$, respectively. (**b**) The schematic circuit diagram of the measurement setup.

4

**SI section-2: Gain calibration:**

We have estimated the gain of amplification chain from temperature dependent Johnson-Nyquist noise (thermal noise)[4]. At zero impinging current equilibrium voltage noise spectrum is given by

$$S_V = g^2(4k_B T R + V_n^2 + i_n^2 R^2) BW \tag{S2}$$

where $g$ is the total gain of amplification chain, $k_B$ the Boltzmann factor and T the temperature, $V_n^2$ and $i_n^2$ are the intrinsic voltage and current noise of the amplifier, and BW is the frequency bandwidth. At an integer quantum Hall plateau, any change in temperature of mixing chamber (MC) plate will only affect the first term in Eq. (S2), while all other terms are independent of temperature. If one plot the $\frac{S_V}{BW}$ as a function of temperature, the slope of the linear curve will be equal to $4g^2 k_B R$. Since at quantum Hall plateau resistance R is exactly known, one can easily calculate the gain of the amplification chain and from the intercept, intrinsics noise of amplifier can be found. So the gain $g$ can be calculated using following equation

$$g = \sqrt{\left(\frac{\partial \left(\frac{S_V}{BW}\right)}{\partial T}\right)\left(\frac{1}{4k_B R}\right)} \tag{S3}$$

In Eq. (S3), $\left(\frac{\partial \left(\frac{S_V}{BW}\right)}{\partial T}\right)$ is the slope of the linear fit.

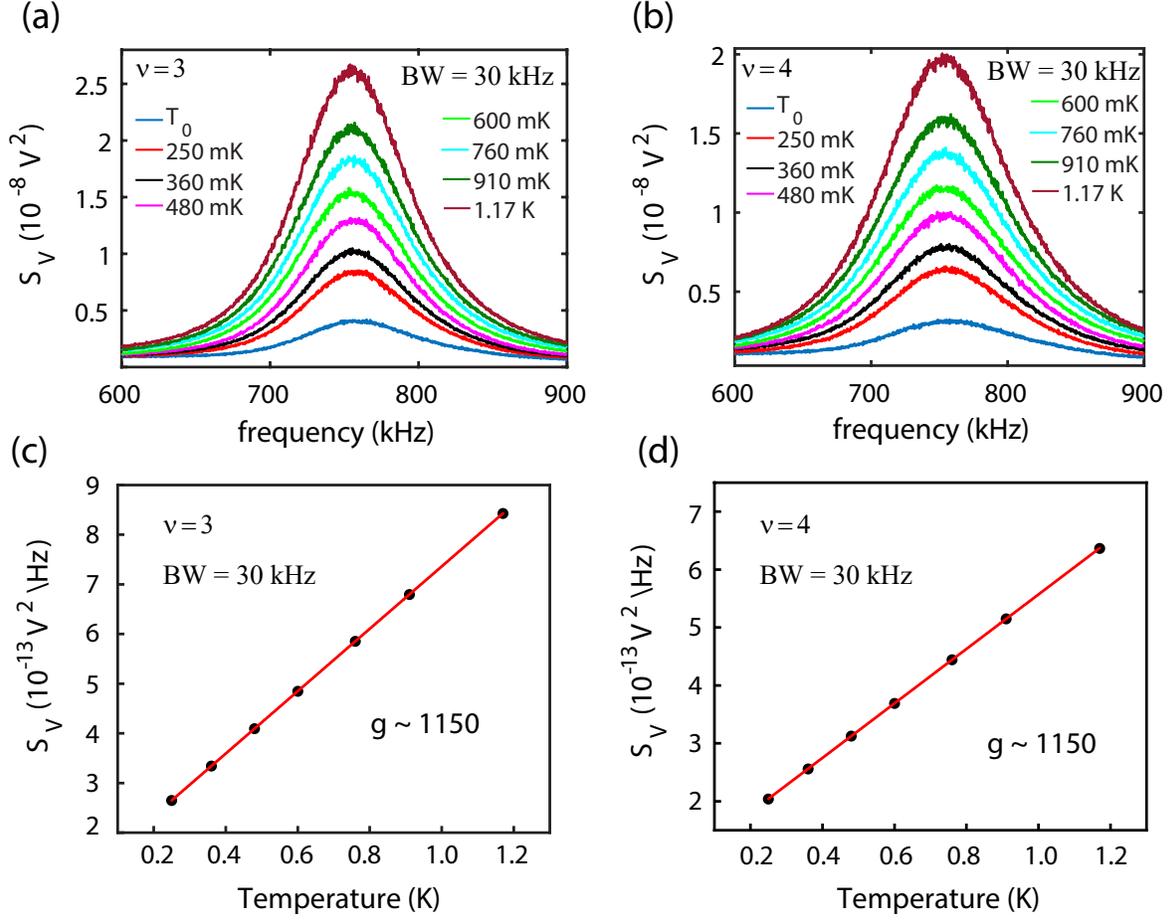

**Fig. S 3: Gain of the amplification chain.** (a) Noise measured at zero bias by spectrum analyzer is plotted as a function of frequency at different temperature at $\nu = 3$ and similarly for (b) $\nu = 4$. From these plots, resonance frequency of tank circuit was found to be ∼760 kHZ. The blue curve in both plot shows the noise data for the base temperature of mixing chamber plate. (c) Symbols represent the plot of noise divided by bandwidth (BW) at resonance frequency as a function of temperature at $\nu = 3$. Solid red line is the linear fit of data. Using Eq. (S3) and slope information of this linear fit, calculated gain was found to be equal to ∼1150. (d) Symbols represent the plot of noise divided by bandwidth (BW) at resonance frequency as a function of temperature at $\nu = 4$. Solid red line is the linear fit of data and from the slope of this line, calculated gain was found to be equal to ∼1150. From the intercept of these linear fit of $\nu = 3$ and 4, the $V_n$ and $i_n$ of amplifier were found to be $\sim 255 \frac{pV}{\sqrt{Hz}}$ and $\sim 22 \frac{fA}{\sqrt{Hz}}$, respectively. The extracted gain from both filling factors are same as expected.

### SI section-3: Electron temperature ($T_0$) determination:

As mentioned earlier in SI section-2, that noise measured by spectrum analyser at zero bias is given by

$$S_V = g^2(4k_BTR + V_n^2 + i_n^2 R^2)BW \tag{S4}$$

As we plot the $\frac{S_V}{BW}$ as a function of temperature at fixed quantum Hall filling as shown in SI figure 3c and 3d, the linear fit of the data will give an linear equation which relates the $\frac{S_V}{BW}$ to temperature T directly. Since we have already estimated the gain (from the slope) and intrinsic noise of amplification chain (from the intercept) (see caption of Fig. S3), corresponding electron temperature $T_0$ at base temperature of mixing chamber plate can be found directly from the known value of measured noise at zero bias. $T_0$ is given by

$$T_0 = \frac{\left(\left(\frac{S_V}{g^2 BW}\right) - (V_n^2 + i_n^2 R^2)\right)}{4k_B R} \tag{S5}$$

As our measured value of noise at base temperature were $16.75 \times 10^{-8} V^2$ at $\nu = 3$ which corresponds to $T_0 \sim 30$ mK. Similarly, at $\nu = 4$ measured noise was $13.34 \times 10^{-8} V^2$, which also corresponds to $T_0 \sim 30$ mK. Similarly, we have estimated the electron temperature at measurement of device-2. While performing the noise measurement on D2 device, the extracted gain of amplification chain was found $\sim 1200$. Following the similar approach as discussed above, electron temperature at base temperature was found to be $\sim 39$ mK. Note that, in our previous work[5] we had measured the electron temperature of the same dilution fridge by measuring the shot noise at a quantum Hall beam splitter and the values were within $\sim 30 - 40$ mK.

### SI section-4: Joule heating and temperature ($T_M$) of the floating reservoir:

The floating reservoir reaches a new equilibrium potential $V_M = \frac{I_S}{2\nu G_0}$ with the filling factor $\nu$ of graphene determined by the $V_{BG}$, whereas the potential of the $S$ contact is $V_S = \frac{I_S}{\nu G_0}$. Thus, the power input to the floating reservoir is $P_{in} = \frac{1}{2}(I_S V_S) = \frac{I_S^2}{2\nu G_0}$, where the pre-factor of $\frac{1}{2}$ results due to the fact that equal power dissipates at the source and the floating reservoirs in Fig. 1a. Similarly, the outgoing power from the floating reservoir is $P_{out} = \frac{1}{2}(2 \times \frac{I_S}{2} V_M) = \frac{I_S^2}{4\nu G_0}$. Thus, the resultant injected power dissipation in the floating reservoir due to joule heating is $J_Q = P_{in} - P_{out} = \frac{I_S^2}{4\nu G_0}$. Another alternative way to quantify the dissipation in floating contact is to calculate the power dissipation at hot spots[6]. Whenever there will be change in chemical potential near contacts, hot spots will generate heat. There are two hot spot near the floating contact. Power dissipated at hot spot to the left of floating contact is given by $P_L = \int_0^{V/2} I dV$. Since $I = V\nu G_0$, hence $P_L$ becomes $P_L = \int_0^{V/2} \left(\nu G_0 V\right) dV = \frac{\nu G_0}{2}\left(\frac{V^2}{4}\right)$. Similarly, power dissipation at the right side hot spot will be $P_R = \frac{\nu G_0}{2}\left(\frac{V^2}{4}\right)$. Hence, total power dissipation at floating contact will be $J_Q = P_L + P_R = \frac{\nu G_0 V^2}{4}$ or in terms of DC current $I_S$ dissipated power will be $J_Q = \nu G_0 \frac{(I_S/\nu G_0)^2}{4} = \frac{I_S^2}{4\nu G_0}$.

The resulting increase in the electron temperature $(T_M - T_0)$ of floating reservoir is determined from the excess thermal noise [7–11]: $S_I = 2G^* k_B (T_M - T_0)$ with $\frac{1}{G^*} = \frac{1}{G_L} + \frac{1}{G_R}$, where $G_L$ and $G_R$ are the conductance of left and right channel respectively. So in our device structure, $\frac{1}{G^*} = \frac{1}{\nu G_0} + \frac{1}{\nu G_0}$, hence $S_I = \nu k_B (T_M - T_0) G_0$.

**SI section-5: Source noise from the contact resistance.**

If the injection contact $(S)$ is not transparent for the edge channels, it will give extra noise at finite DC bias. This unwanted noise is known as source noise whose magnitude will be given as $2eI(1-t)$, where $t$ is the transmittance of source contact. To estimate the noise generated by the source contact itself, we have extracted out the contact resistance of the source contact following a measurement circuit shown in Fig. S4.

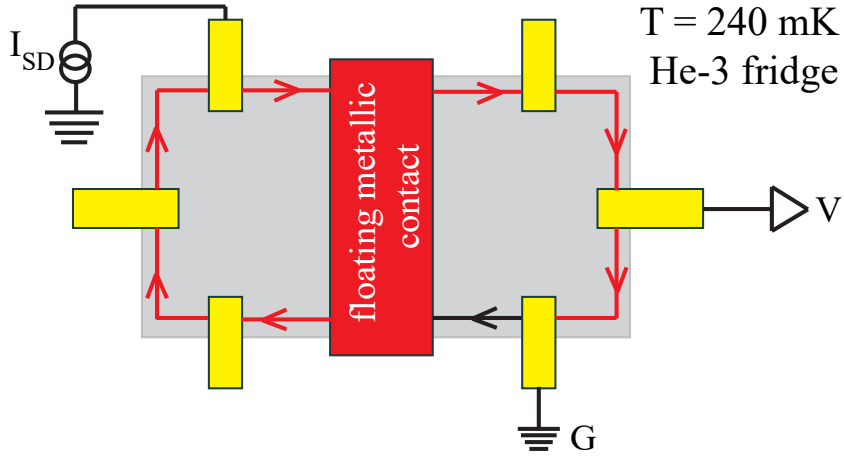

**Fig. S 4: Source noise from the contact resistance:** To determine the contact resistance of the source contact, three probe measurement technique was used. For our device structure, voltage probe $V$ situated in path of hot edge for given chirality will measure a voltage of $V_H = I(R_0 + R_L + R_C)$ where $R_0$ is the quantum Hall resistance, $R_C$ - contact resistance and $R_L$ - line resistance of ground contact (here the ground was at room temperature). On the other hand for opposite chirality (by changing the polarity of the magnetic field), it will measure a voltage of $V_C = IR_L$. Differences of these two measured voltages for opposite charities will give us $I(R_0+R_C)$. Since in quantum Hall, at plateau, conductance is just the inverse of measured resistance, so conductance will be $\frac{1}{R_0+R_C}$. Hence transmittance $t$ will be given by the ratio of actual quantum resistance at that plateau and measured resistance, i.e. $t = \frac{R_0}{R_0+R_C}$. Once the transmittance is known, one can easily estimate the source noise $2eI(1-t)$. Since amplifier is situated in path of reflected current, it will measure only part of the generated source noise. In particular, for our device structure, only two open arms are available for current flow, amplifier will always see only the half of total no of edge channels leaving the floating reservoir. So it will always measure the only $\frac{1}{4}$th of the source noise generated at source contact. Values of the contact resistance, transmittance and maximum source noise for D1 is given in Table S1.

9

| Filling Factor($\nu$) | Hot edge $V_H/I = (R_0 + R_L + R_C)(\Omega)$ | Cold edge $V_C/I = R_L$ ($\Omega$) | Contact Resistance $(V_H/I - V_C/I) - R_0$ in ($\Omega$) | Trans- mittance (t) | Source Noise / 4 ($10^{-29} A^2$/ Hz) at $I_{max}$ |
|---|---|---|---|---|---|
| 1 | 26485 | 632 | 40 | 0.9984 | 0.0256 @ $I_{max}$ = 2.0 nA |
| 5/3 | 16157 | 632 | 38 | 0.9976 | 0.0384 @ $I_{max}$ = 2.0 nA |
| 2 | 13572 | 632 | 34 | 0.9974 | 0.0416 @ $I_{max}$ = 2.0 nA |
| 7/3 | 11732 | 632 | 38 | 0.9966 | 0.0544 @ $I_{max}$ = 2.0 nA |
| 8/3 | 10348 | 632 | 36 | 0.9963 | 0.0592 @ $I_{max}$ = 2.0 nA |
| 3 | 9271 | 632 | 35 | 0.9959 | 0.0656 @ $I_{max}$ = 2.0 nA |

**Table S1.: Contact resistance and the Source noise.** From the transmittance column of above table, it is clear that reflection is always less than 1% for all measured filling factor. So the measured source noise could be as maximum as about $0.06 \times 10^{-29} A^2 Hz^{-1}$ at maximum source current. This source noise is $\sim 70$ times smaller than the measured excess thermal noise. Line resistances of the He-3 fridge was measured separately and agrees with $632\Omega$.

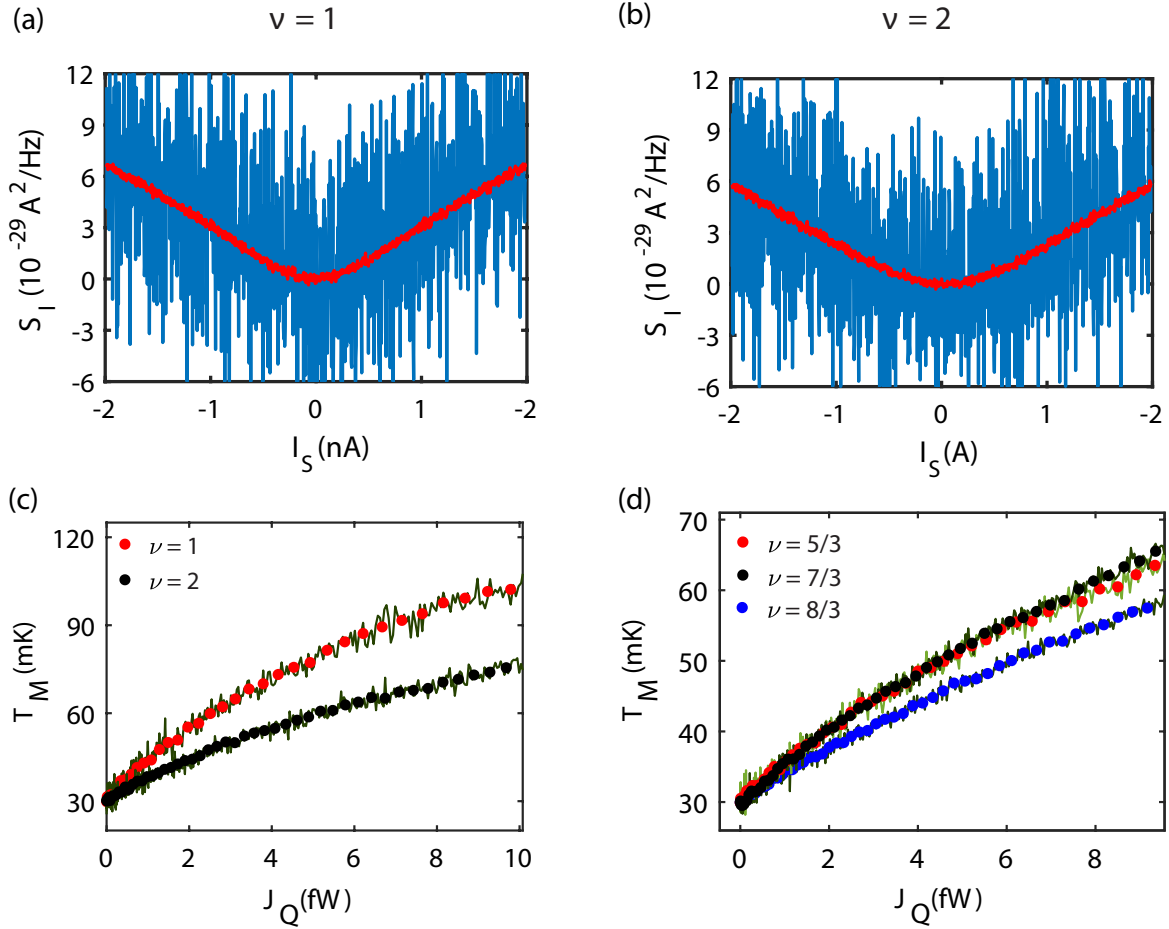

**Fig. S 5: Extended excess thermal noise data(a,b)** Excess noise for single scan (blue) and the average of 1000 scan (red) for $\nu = 1$ and 2 of D1 device. (**c**) Solid curves show the data extracted directly from raw excess thermal noise data for D1 device for $\nu =1$ and 2, respectively. Solid circles display the 9 point average of corresponding raw data which is shown in the manuscript (Fig. 2b). (**d**) Solid curves data extracted from raw excess thermal noise data for D1 device for $\nu =$5/3, 7/3 and 8/3, respectively. Solid circles display the 9 point average of corresponding raw data which is shown in the manuscript (Fig. 3c)

11

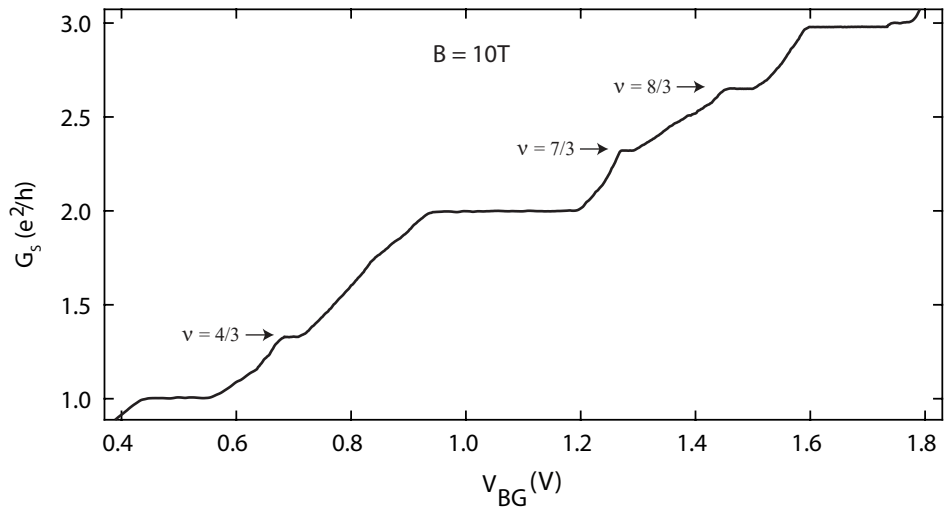

**Fig. S 6: Quantum Hall response of D2 device.** Gate response of D2 device has been plotted as a function of the back gate voltage at 10T. In this plot conductance is measured at the source contact as shown in Fig. 1c. The fractional plateaus are marked by arrowheads. In this device also, measured conductance are consistent with the charge equilibration of counter propagating modes.

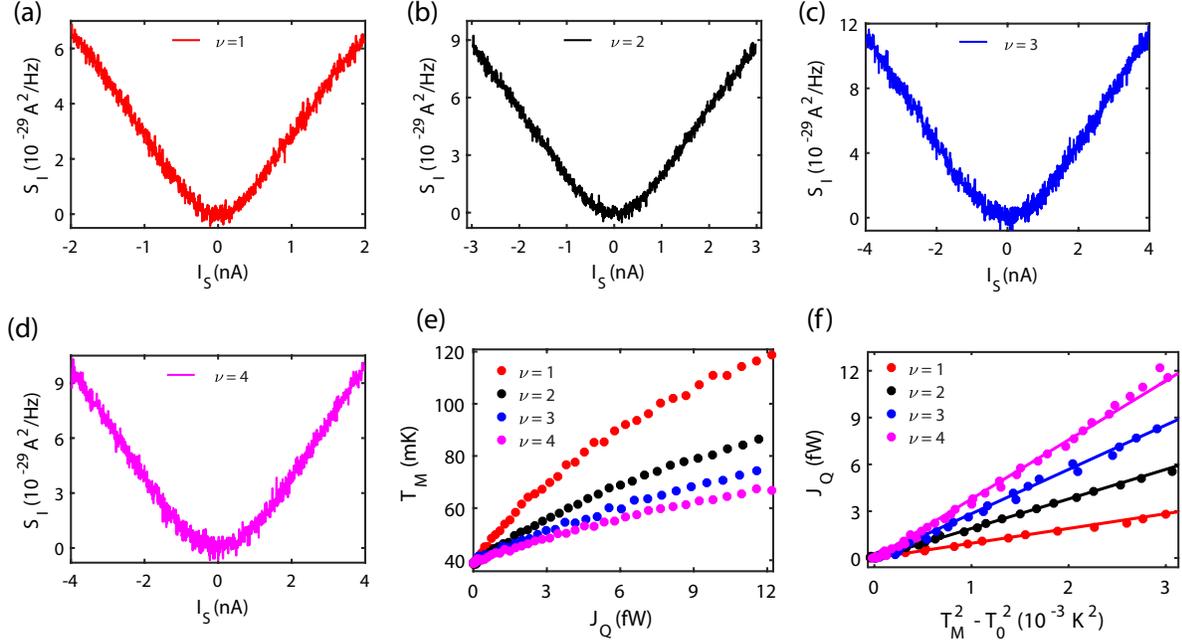

**Fig. S 7: Extended data of D2 device (Integer quantum Hall thermal conductance)** (**a**) Excess thermal noise $S_I$ is plotted as a function of source current $I_S$ at filling $\nu = 1$ (a), 2 (b), 3 (c), and 4 (d). (**e**) Temperature $T_M$ (extracted from the excess thermal noise shown in (a,b,c,d)) of floating contact is plotted as a function of dissipated power $J_Q$ (obtained using $J_Q = \frac{I_S^2}{4\nu G_0}$) for filling factors $\nu = 1$ (red), 2 (black), 3 (blue), and 4 (magenta), respectively. Symbols display the extracted temperature data using equation $S_I = \nu k_B (T_M - T_0) G_0$. (**f**) The $J_Q$ is plotted as a function of $T_M^2 - T_0^2$. Solid circles display the data and the solid lines are the theoretically predicted contribution of electronic thermal conductance at these filling factors. Slope of linear fits of theses data gives us the thermal conductance values 0.99, 2.05, 3.04 and 3.96 $\kappa_0 T$ for $\nu = 1, 2, 3$ and 4, respectively. This measured value again support the accuracy of our measurement set-up.

13

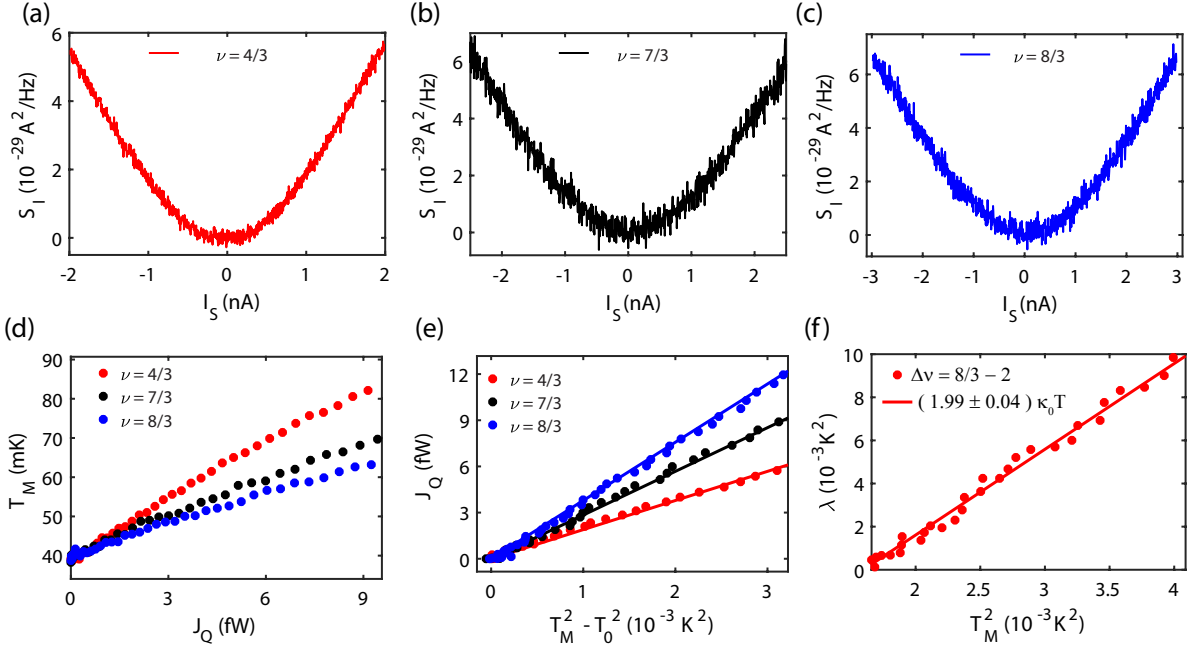

**Fig. S 8: Extended data of D2 device (Fractional quantum Hall thermal conductance)** Excess thermal noise $S_I$ is plotted as a function of source current $I_S$ at $\nu$ = 4/3 (a), 7/3 (b) and 8/3 (c). **(d)** The increased temperatures $T_M$ of the floating reservoir are plotted (solid circles) as a function of dissipated power $J_Q$ for $\nu$ = 4/3 (red), 7/3 (black) and 8/3 (blue), respectively. **(e)** $J_Q$ is plotted (solid circles) as a function of $T_M^2 - T_0^2$ for $\nu$ = 4/3 (red), 7/3 (black) and 8/3 (blue), respectively. The solid red, black and blue lines represent the theoretical contribution to electronic heat flow with thermal conductance of $2\kappa_0 T$, $3\kappa_0 T$, and $4\kappa_0 T$, respectively. **(f)** The $\lambda = \Delta J_Q/(0.5\kappa_0)$ is plotted as a function of $T_M^2$ for $\Delta\nu = 8/3 - 2$, where $\Delta J_Q = J_Q(\nu_i, T_M) - J_Q(\nu_j, T_M)$. The solid lines are the fittings to extract the thermal conductance of 2/3 like fractional quantum Hall state. Extracted thermal conductance was found to be $1.99 \pm 0.04 \kappa_0 T$ for $\Delta\nu = 8/3 - 2$.

14

**SI section-6: Longitudinal resistance of D3 device.**

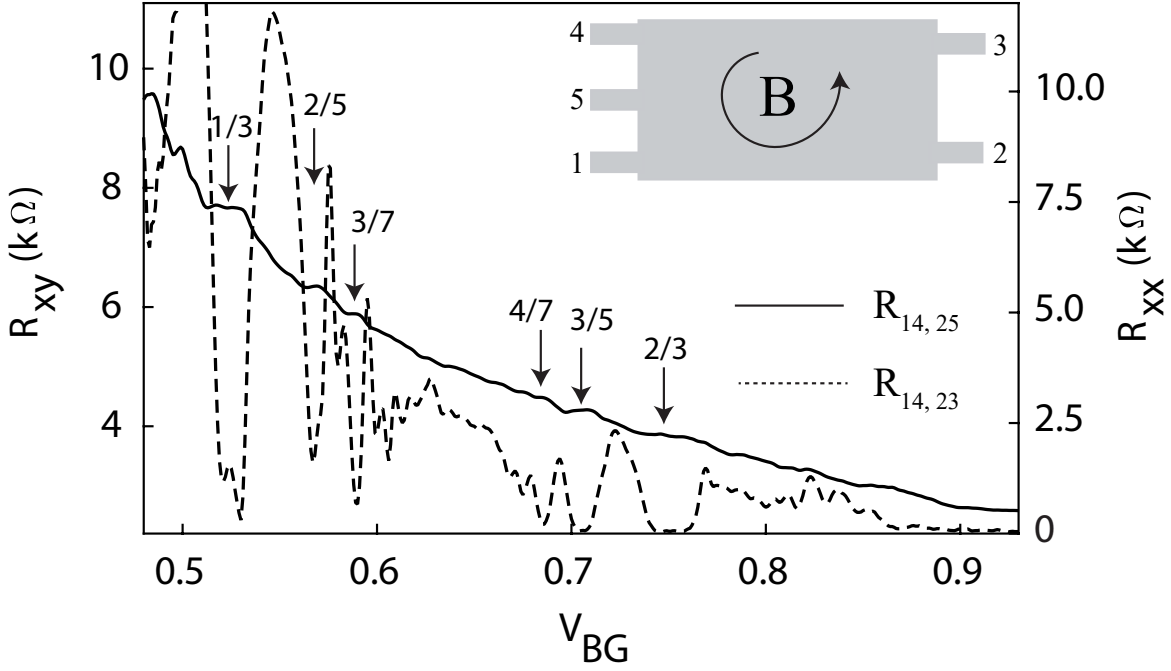

**Fig. S 9: Longitudinal resistance of D3 device.** Gate response of D3 device has been plotted as a function of the botoom graphite gate voltage at 10T. The solid trace represents the measured Hall resistance ($R_{xy}$), while dashed curve corresponds to measured longitudinal resistance ($R_{xx}$). One can clearly observe the plateau at particle like filling factor $\nu$ = 1/3, 2/5, and 3/7 accompanied by minima in $R_{xx}$. The plateaus at hole conjugate states $\nu$ = 2/3, 3/5, and 4/7 are also well developed accompanied by the minima in $R_{xx}$, which approaches to zero. The plateaus are marked by arrowheads. Inset shows the configuration of contacts in device-3. $R_{xy}$ is measured by passing the current between contacts 1 and 4, and voltage is measured between 2 and 5, while for $R_{xx}$ measurement, 2 and 3 are used as voltage probes.

**SI section-7: Thermal conductance contribution from the bulk.**

It can be seen from Fig. S9 that $R_{xx}$ for D3 device goes to zero for hole-like FQH. Though, we have not measured the $R_{xx}$ for D1 and D2 devices due to limited contact configurations, but the flatness of the plateaus in Fig. 1c and Fig. S6 indicate that the $R_{xx}$ is close to zero in those devices. Note that the contact resistance values in table 1 contains the actual contact resistance as well as $R_{xx}$ component, and can be seen that it has similar values within $\sim 35 - 40\Omega$ for both integer and FQH states. The integer plateaus are so robust in graphene (Fig. 1c and Fig. S6) that zero $R_{xx}$ is inevitable as seen for D3 device and thus,

15

table 1 indicates that $R_{xx}$ for FQH states in D1 and D2 devices are close to zero. Furthermore, we note that the heat conduction through the bulk would influence the results for all fractions including those without counter propagating modes (integers or 4/3 or 7/3). So, the fact that the values there are as expected is an experimental argument in favor of negligible contribution of bulk heat conductance in our devices.

**SI section-8: Values of electrical conductance without charge equilibration.**

To estimate the value of electrical conductance of hole like fractional quantum Hall state without charge equilibration along the propagation length for our device configuration, we follow the approach of Landauer Büttiker model[12]. Note that in this calculation we assume full equilibration at the contacts including the floating reservoir. We will calculate the electrical conductance of 2/3, 5/3 and 8/3 state, which host one counter propagating edge state of conductance 1/3. The schematic of the device with contact number is shown in Fig. S10.

we assume the number of charge mode with charge $e$ is N. For multi-probe device, the net current flowing in $i^{th}$ contact is given by

$$I_i = \sum_j (G_{j \leftarrow i} V_i - G_{i \leftarrow j} V_j) \tag{S6}$$

where $G_j \leftarrow i$ is the conductance from $i^{th}$ contact to $j^{th}$ contact and $V_i$ is the voltage of $i^{th}$ contact.

$$\begin{pmatrix} I_1 \\ I_2 \\ I_3 \\ I_4 \\ I_5 \\ I_6 \\ I_7 \end{pmatrix} = \frac{e^2}{h} \begin{pmatrix} N+1/3 & -1/3 & 0 & 0 & 0 & 0 & -N \\ -N & 2(N+1/3) & -1/3 & 0 & -N & -1/3 & 0 \\ 0 & -N & N+1/3 & -1/3 & 0 & 0 & 0 \\ 0 & 0 & -N & N+1/3 & -1/3 & 0 & 0 \\ 0 & -1/3 & 0 & -N & N+1/3 & 0 & 0 \\ 0 & -N & 0 & 0 & 0 & N+1/3 & -1/3 \\ -1/3 & -0 & 0 & 0 & 0 & -N & N+1/3 \end{pmatrix} \begin{pmatrix} V_1 \\ V_2 \\ V_3 \\ V_4 \\ V_5 \\ V_6 \\ V_7 \end{pmatrix} \tag{S7}$$

Since contact $4^{th}$ and $7^{th}$ are grounded. So we can eliminate those from above matrix Eq. (S7). After eliminating the rows and column associated with contacts no 4 and 7, we get

16

$$\begin{pmatrix} I_1 \\ I_2 \\ I_3 \\ I_5 \\ I_6 \end{pmatrix} = \frac{e^2}{h} \begin{pmatrix} N+1/3 & -1/3 & 0 & 0 & 0 \\ -N & 2(N+1/3) & -1/3 & -N & -1/3 \\ 0 & -N & N+1/3 & 0 & 0 \\ 0 & -1/3 & 0 & N+1/3 & 0 \\ 0 & -N & 0 & 0 & N+1/3 \end{pmatrix} \begin{pmatrix} V_1 \\ V_2 \\ V_3 \\ V_5 \\ V_6 \end{pmatrix} \quad (S8)$$

For $\nu = 5/3$, N=2, then Eq. (S8) becomes

$$\begin{pmatrix} I_1 \\ I_2 \\ I_3 \\ I_5 \\ I_6 \end{pmatrix} = \frac{e^2}{h} \begin{pmatrix} 2+1/3 & -1/3 & 0 & 0 & 0 \\ -2 & 2(2+1/3) & -1/3 & -2 & -1/3 \\ 0 & -2 & 2+1/3 & 0 & 0 \\ 0 & -1/3 & 0 & 2+1/3 & 0 \\ 0 & -2 & 0 & 0 & 2+1/3 \end{pmatrix} \begin{pmatrix} V_1 \\ V_2 \\ V_3 \\ V_5 \\ V_6 \end{pmatrix} \quad (S9)$$

Since current is injected at contact 1. so current column becomes

$$\begin{pmatrix} I_1 \\ I_2 \\ I_3 \\ I_5 \\ I_6 \end{pmatrix} = \begin{pmatrix} I \\ 0 \\ 0 \\ 0 \\ 0 \end{pmatrix} \quad (S10)$$

hence voltage will be

$$\begin{pmatrix} V_1 \\ V_2 \\ V_3 \\ V_5 \\ V_6 \end{pmatrix} = I \frac{h}{e^2} \begin{pmatrix} 0.4633 \\ 0.2432 \\ 0.2085 \\ 0.0347 \\ 0.2085 \end{pmatrix} \quad (S11)$$

so voltage measured at contact 1 will be

$$V_1 = 0.4633 \times I \frac{h}{e^2} \quad (S12)$$

so conductance at source contact 1 at 5/3 filling factor will be

$$G = \frac{I}{V_1} = 2.16 \frac{e^2}{h} \quad (S13)$$

Similarly for 8/3 filling factor, calculated conductance will be

$$G = \frac{I}{V_1} = 3.16 \frac{e^2}{h} \quad (S14)$$

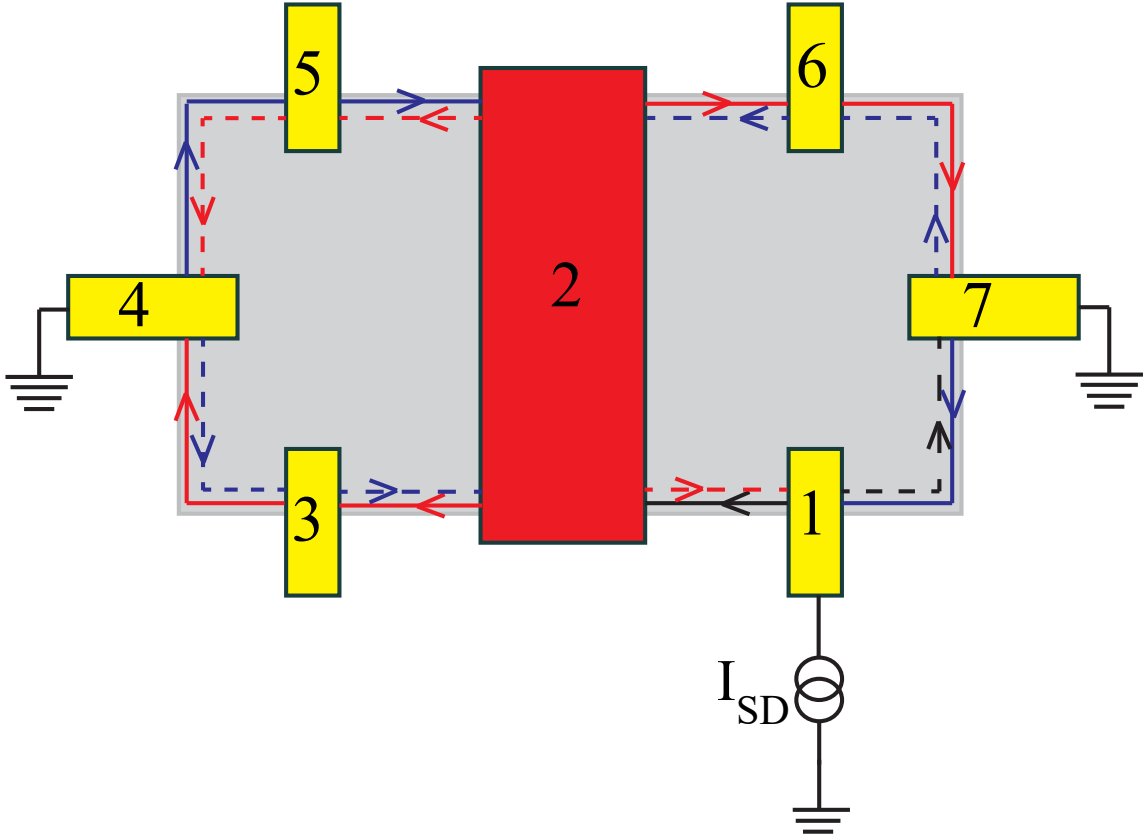

**Fig. S 10: Schematic for the electrical conductance calculation using Landauer Büttiker Formalism.** The contacts are marked by the numbers. The charged mode with charge $e$ is shown by solid line and charge mode with charge $e/3$ is shown by dashed line.

These are summarized in table 2. Calculated value for the $5/3$ filling factor was found to be $2.16\frac{e^2}{h}$ and $4.80\frac{e^2}{h}$ at source(S) and reflected/transmitted (R/T) contacts, respectively. Similarly, for the $8/3$ filling factor, calculated conductance was found to be $3.16\frac{e^2}{h}$ and $6.75\frac{e^2}{h}$ at source(S) and reflected/transmitted (R/T) contacts, respectively. However, in our experiment, the measured values of conductance at source ($I_S/V_S$) and reflected/transmitted ($I_S/V_R$ or $I_S/V_T$) contacts were found to be $1.67\frac{e^2}{h}$ and $3.33\frac{e^2}{h}$ for $5/3$ filling factor, and $2.67\frac{e^2}{h}$ and $5.33\frac{e^2}{h}$ for $8/3$ filling factor, respectively. These measured values suggest that the charge equilibration of the counter propagating edge modes along the propagation length is well established in our devices. This was further verified in D3 device via 2-probe conductance measurement as described in the main manuscript.

| Filling Factor($\nu$) | Calculated conductance in absence of charge equilibration (I/V$_i$) (in e²/h ) | | | Experimentaly measured conductance (I/V$_i$) (in e²/h ) | | |
|---|---|---|---|---|---|---|
| | Source (1) | Reflected (2) | Transmitted (3) | Source (1) | Reflected (2) | Transmitted (3) |
| 5/3 | 2.16 | 4.80 | 4.80 | 1.67 | 3.33 | 3.33 |
| 8/3 | 3.16 | 6.75 | 6.75 | 2.67 | 5.33 | 5.33 |

**Table S2: Comparison of calculated electrical conductance using Landauer Büttiker Formalism and measured conductance.** Electrical conductance calculated for the hole like states assuming no charge equilibration between the counter propagating edges is always much higher than the experimentally measured values.

**SI section-9: Heat loss by electron-phonon Cooling.**

The heat balance equation mentioned in main manuscript, in addition to the electronic contribution, there is another mechanism of heat transfer via electron-phonon cooling ($J_Q^{e-ph}$). To estimate the contribution of $J_Q^{e-ph}$, we have subtracted the electronic contribution ($J_Q^e$) from the total dissipated power ($J_Q$),

$$J_Q^{e-ph} = J_Q - J_Q^e \tag{S15}$$

Since $J_Q^e = 0.5N\kappa_0(T_M^2 - T_0^2)$, where $N$ is toal number of electronic channel, leaving the floating contact. Usually, $J_Q^{e-ph}$ has the functional form of $J_Q^{e-ph} = \beta(T_M^q - T_0^q)$. In our devices, $J_Q^{e-ph}$ was found to be negligible. We observed that it only has significant effect for filling factor 1, in device D1, but after the electron temperature of $\sim 60 - 80$ mK. The power exponent ($q$) was found to be 5 in this case. It should be worth to mention here that $q$ was found to be varying from 4-6 in our earlier observation[5] and reported elsewhere also[13].

19

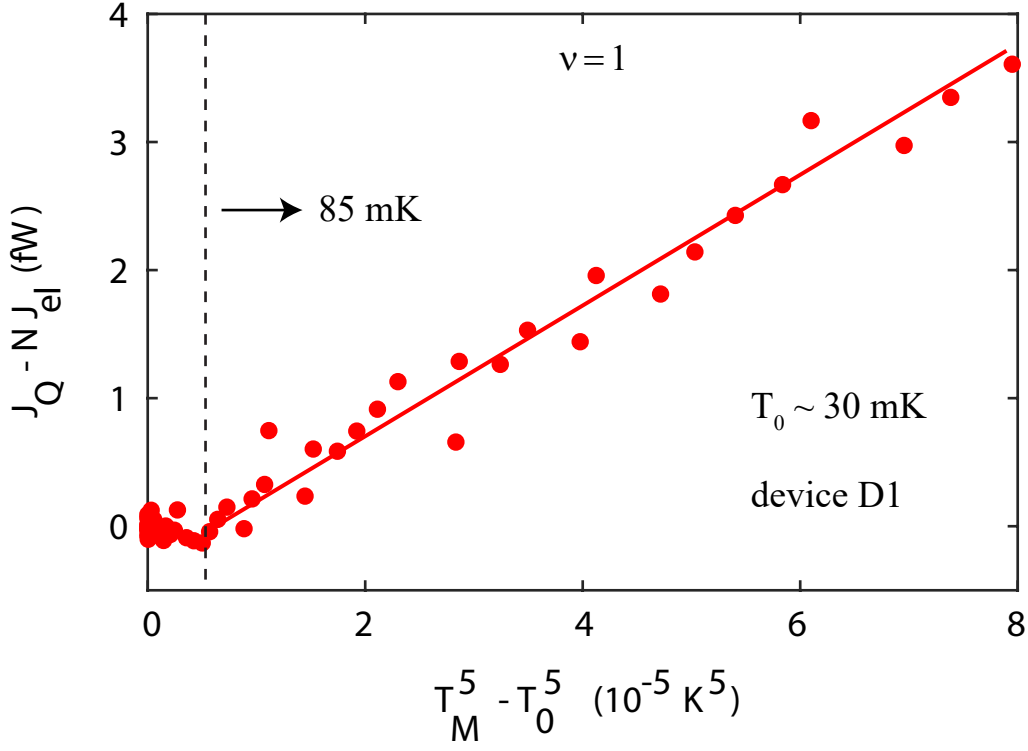

**Fig. S 11: Heat loss by electron-phonon coupling.** Solid circles display the electron-phonon contribution of the heat loss ($J_Q^{e-ph}$) as a function of $(T_M^5 - T_0^5)$ for $\nu = 1$ for D1 device. The solid line is the linear fit with a slope, $\beta \sim 0.05 nW/K^5$. The vertical dashed line shows that up to $\sim 85$ mK, $J_Q^{e-ph}$ is very small ( < 0.15 fW), which is negligible compared to electronic part. Here, we have not plotted the heat loss to due to electron-phonon coupling for higher filling factors as it was very small at those fillings.

**SI section-10: Effect of Heat Coulomb Blockade**

While extracting the electronic contribution to thermal conductance for $2/3$ part of $5/3$ and $8/3$ filling factor, we have subtracted the data of filling factor 1 and 2, from $5/3$ and $8/3$, respectively as discussed in main text. This subtraction helps us to remove the electron phonon contribution, although it was found negligible below $\sim 85$ mK, as discussed in Section S8. But other non-trivial effect, which might play an important role in extracting the thermal conductance without subtraction is the heat Coulomb blockade. It was predicted theoretically[14] and shown experimentally[13] that this effect will cause a reduction in measured thermal conductance, if the electron temperature of floating reservoir is significantly smaller than the crossover

characteristic temperature ($T_{CB}$). $T_{CB}$ is directly related to the capacitance (C) of floating metallic reservoir

$$T_{CB} = \frac{\nu e^2}{\pi k_B C} \tag{S16}$$

For our device D1 and D2, using the hBN thickness of $\sim$ 10nm ( between the floating reservoir and bottom graphite estimated from the rate of reactive ion etching), and area of floating contact of 7 $\mu m^2$, the geometrical capacitance of floating contact was found to be $\sim$ 25 fF. This will correspond to a crossover temperature of $\sim$ 24 mK, which is even smaller than our minimum electron temperature of $\sim$ 30 mK. This suggest that the effect of heat Coulomb blockade will be negligible in our devices.

**SI section-11: Universality of measured thermal conductance**

Quantum Hall phases in bilayer graphene (BLG)/mono layer graphene (MLG) are much richer than the conventional GaAs/AlGaAs based system. The presence of an extra degree of freedoms in BLG, which includes valley and orbital degree of freedom in addition to the spin, makes the quantum Hall phases exceptionally tunable with external experimental parameters. For example, when all symmetries are broken at a high magnetic field[15,16], different quantum Hall states belong to different orbital degrees of freedom, named N=0 and N=1 landau levels. In our case, we have performed the experiments on two devices, which are in the opposite regime, namely p-doped for the device D1 and n-doped for the device D2. In Device D1, the observed fractional quantum Hall states are 5/3, 7/3, and 8/3. As shown in the main manuscript, the thermal conductance of both the 5/3 and 8/3 states suggests that the counter propagating modes of the 2/3 level are thermally non-equilibrated, although their wave-functions belong to the different orbital degree of freedom, named N=0 and N=1, respectively[15]. We also observe that the measured thermal conductance of 7/3 and 8/3 states in hole doped regime (device D1) and in electron doped regime (device D2) are the same, irrespective of the different orbital nature of their wave-functions, which is N=1(0) in hole(electron) doped regime[15]. These observations suggest that the measured thermal conductance of fractional quantum Hall states are universal, and they do not depend on the orbital or valley degree of freedom of Landau levels. In Fig. S 12, the orbital degree of freedom in different filling ranges are shown for electron and hole-doped regimes. Since we observe similar behaviour for different orbital indices in BLG, we also believe that the measured thermal conductance of fractional quantum Hall states will show the similar behaviour in monolayer graphene irrespective of the orbital nature of Landau levels. This claim is supported by the measured thermal conductance of 4/3 state, which is found to be $2\kappa_0 T$ in device D2 and in our earlier experiment on monolayer graphene[5].

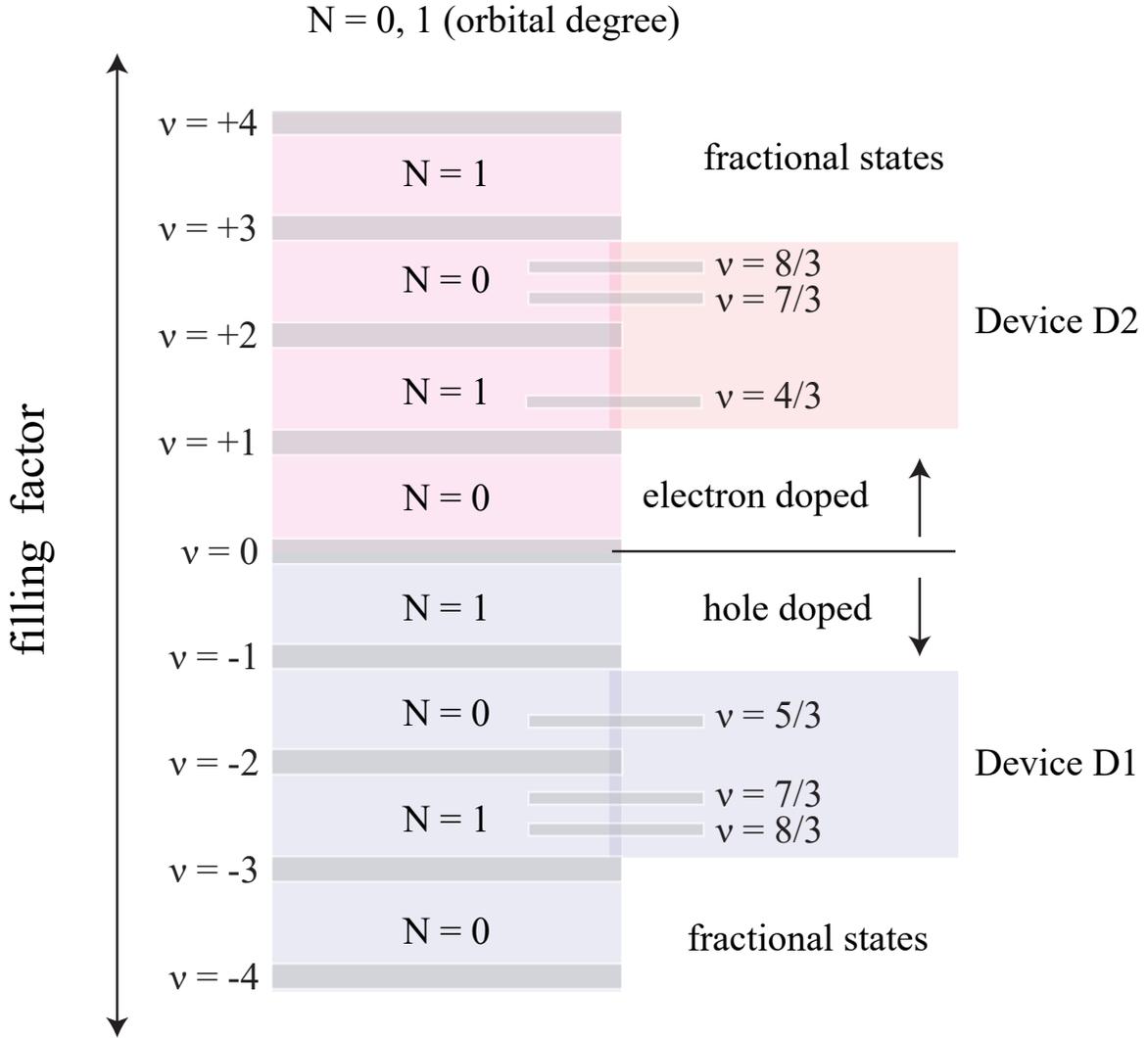

**Fig. S 12: Quantum Hall states and associated orbital degree of freedom.** In bilayer graphene, when the symmetries are not broken, $N = 0$ and $N = 1$ landau levels are degenerate. Due to spontaneous symmetry breaking at a high magnetic field, one observes the plateaus in conductance at $1e^2/h, 2e^2/h, 3e^2/h,$ and $4e^2/h$ in the hole and electron-doped regime. In a high mobility sample, when the landau levels are partially filled, one observes the fractional quantum Hall states also. Although the conductance of the fractional quantum Hall states is the same for electron-doped and hole-doped regime both, the orbital degree of freedom of these states' wavefunction is not the same. In the above figure, the orbital degree of freedom (N=0 or 1) is marked for electron and hole-doped regimes for different filling ranges[15,16]. Observed fractional states of devices D1 and D2 are also marked.

**SI section-12: Effect of Displacement field on measured thermal conductance**

A perpendicular displacement field applied to BLG breaks the inversion symmetry of BLG and hence opens a gap at the charge neutrality point. At a very high magnetic field, due to the spontaneous symmetry breaking, different quantum Hall states emerge with a complex degree of freedom. In BLG, interplay of the displacement field and magnetic field causes the transition between different kind of valley and spin degree of freedom in integer and fractional quantum Hall state[17]. In our experiments, measurements were performed at some finite displacement field only. We can only tune the displacement field and carrier density independently if both top and bottom gate is present. Unfortunately, we could not measure the thermal conductance in a dual gated device, because of the presence of the central, floating contact, which acts as a hot reservoir, upon injecting a DC current. In the presence of a top gate, this floating contact will directly short with the top gate electrode. While we did not perform our measurements exactly at zero electric field, we expect that the quantization of the thermal transport can be affected by the displacement field at some particular values due to the following reason. It has been shown that integer as well as fractional quantum Hall plateaus undergo phase transitions at zero and some particular displacement field (which depends on the filling factor)[17]. This is clearly visible also in electrical conductance measurements[17]. Since the quantization of the thermal conductance as well as electrical conductance at a fixed filling factor are related to the same topological state, we expect a similar transition in the thermal conductance also at zero and some particular displacement field (depends on filling factor).

## SI section-13: Theoretical calculations of equilibration lengths

In this section, we derive charge and heat equilibration lengths for transport on an edge with counter-propagating 1 and 1/3 modes. First, we compute charge and energy currents and conductances induced by inter-channel tunneling. We then relate these conductances to characteristic length scales for charge and heat equilibration. Our calculations reveal that strong interactions can cause parametrically different length scales for charge and heat equilibration, which in accordance with our experimental observations.

## The chiral Luttinger liquid description of the FQH edge

To fix the notation, we start by considering a general fractional quantum Hall (FQH) edge. The general chiral Luttinger liquid ($\chi$LL) action reads [18–20]

$$S_{\chi LL} = \frac{1}{4\pi} \int dt \int dx \sum_{ij} \partial_x \phi_i \left( K_{ij} \partial_t - V_{ij} \partial_x \right) \phi_j, \tag{S17}$$

where $\phi_i$ with $i = 1, 2, \ldots, N$, are chiral bosonic fields obeying the equal time commutation relations

$$[\phi_i(x), \phi_j(x')] = i\pi K_{ij}^{-1} \text{sgn}(x - x'). \tag{S18}$$

The quantities $K_{ij}$ are integer elements of a quadratic matrix $K$ encoding the topological properties of the FQH state. More precisely, the $K$-matrix contains information about the bulk filling factor $\nu$, the quasiparticle charges $Q$, and the quasiparticle statistics angles $\Theta$ according to

$$\nu = \sum_{ij} t_i K_{ij}^{-1} t_j, \tag{S19a}$$

$$Q = \sum_{ij} m_i K_{ij}^{-1} t_j, \tag{S19b}$$

$$\frac{\Theta}{\pi} = \sum_{ij} m_i K_{ij}^{-1} m_j. \tag{S19c}$$

In these expressions, the integer numbers $t_i$ are elements of the so-called charge vector. This vector reflects a choice of basis for the gauge fields in the underlying Chern-Simons gauge field theory, and fixes the charges of the quasiparticles with respect to an external electromagnetic field. The most used bases are the *symmetric basis*: $t = (1, 1, 1, \ldots)$ and the *hierarchical* basis: $t = (1, 0, 0, \ldots)$. In the following, we stick to the symmetric basis. The creation or destruction of quasiparticles (including the special case of the charge $e$ excitation resembling an electron) is specified by the string of positive integers $m_i$. The operator creating such a local quasiparticle with charge $Q$ at position $(x, \tau)$ is given by

$$O(x, \tau) = \exp\left[ i \sum_{j=1}^{N} m_j \phi_j(x, \tau) \right]. \tag{S20}$$

24

For a given FQH edge theory, the choice of $K$-matrix is not completely unique. Equivalent representations of the same state are related by a transformation matrix $W \in SL(\mathbb{Z})$ (square matrices with integer entries and unit determinants). The transformation rules read

$$K \to K' = W^T K W, \tag{S21a}$$

$$t_i \to t'_i = \sum_j W^T_{ij} t_j, \tag{S21b}$$

$$m_i \to m'_i = \sum_j W^T_{ij} m_j, \tag{S21c}$$

and leave $\nu$, $Q$, and $\Theta$ invariant.

The matrix, $V$, with components $V_{ij}$ in Eq. (S17), parametrizes both the edge mode velocities (diagonal terms $V_{ii}$) as well as short ranged interactions between the modes (off-diagonal terms). Importantly, this matrix is non-universal (it depends, e.g., on the sample confinement potential), but we demand it to be positive definite. This is needed to bound the Hamiltonian from below and make the theory stable.

The action (S17) is quadratic in the bosonic fields, and can therefore be diagonalized by a few formal manipulations [20]:

1. Let $\phi_i = \Lambda_{1ij} \phi'_j$, where $(\Lambda_1^T \Lambda_1)_{ij} = \delta_{ij}$ such that $(\Lambda_1^T K \Lambda_1)_{ij} = \lambda_i \delta_{ij}$. In other words, $K$ is diagonalized by the orthogonal matrix $\Lambda_1$. The eigenvalues of $K$ are denoted $\lambda_i$, but it is useful to also introduce $\nu_i \equiv \lambda_i^{-1}$.

2. Let $\phi''_i = \Lambda_{2ij} \phi'_j$, where $\Lambda_{2ij} = \frac{\delta_{ij}}{\sqrt{|\lambda_i|}}$. Hence, the eigenvalues of $K$ are scaled such that only their signs remain. What remains of $K$ is then its signature matrix, which we call $\eta$.

After these two steps, we obtain the action

$$S = \frac{1}{4\pi} \int dt \int dx \sum_{ij} \partial_x \phi''_i \left( \eta_{ij} \partial_t - V''_{ij} \partial_x \right) \phi''_j, \tag{S22}$$

where $\eta_{ij} = \mathrm{sgn}(\lambda_i) \delta_{ij}$ and $V'' = \Lambda_2^T \Lambda_1^T V \Lambda_1 \Lambda_2$.

3. Finally, let $\phi''_i = \Lambda_{3ij} \tilde{\phi}_j$, where $\Lambda_3$ diagonalizes $V''$: $(\Lambda_3^T V'' \Lambda_3)_{ij} = \tilde{v}_i \delta_{ij}$. At the same time, we require that $\Lambda_3$ preserves the signature: $\Lambda_3^T \eta \Lambda_3 = \eta$. The action is then completely decoupled into freely propagating chiral eigenmodes

25

$$S = \frac{1}{4\pi} \int dt \int dx \sum_i \partial_x \tilde{\phi}_i \left( \eta_{ii} \partial_t - \tilde{v}_i \partial_x \right) \tilde{\phi}_i. \tag{S23}$$

The eigenmode bosons $\tilde{\phi}_i$ obey

$$\left[ \tilde{\phi}_i(x), \tilde{\phi}_i(x') \right] = i\pi \eta_i \mathrm{sgn}(x - x'). \tag{S24}$$

Stability, in the sense of a Hamiltonian bounded from below requires $\tilde{v}_i > 0$ (this ensured by the previous assumption that $V$ is positive definite). The directions of propagation are instead determined by $\eta_{ii} = \mathrm{sgn}(\lambda_i)$. For convenience, we define the total diagonalization matrix $M \equiv \Lambda_1 \Lambda_2 \Lambda_3$ and write out the diagonalization transformations in terms of $M$

$$\eta_{ii} = \left( M^T K M \right)_{ii} = \mathrm{sgn}(\lambda_i), \tag{S25a}$$

$$\tilde{v}_i \equiv \tilde{V}_{ii} = \left( M^T V M \right)_{ii}, \tag{S25b}$$

$$\tilde{\phi}_i = \sum_j M^{-1}_{ij} \phi_j, \tag{S25c}$$

$$\tilde{t}_i = \sum_j M^T_{ij} t_j, \tag{S25d}$$

$$\tilde{m}_i = \sum_j W^T_{ij} m_j. \tag{S25e}$$

Note that the diagonal signature matrix satisfies $\eta = \eta^{-1}$. While the manipulations above are simple in the formal sense, in practice it can be quite difficult to find the matrix $M$. Moreover, it is not unique. A useful scheme for finding clever choices of $M$ is presented in Ref. [21].

The total charge density (in units where $e = 1$) on the edge is defined as

$$\rho(x) \equiv \sum_i t_i \frac{\partial_x \phi_i(x)}{2\pi} = \sum_i \frac{\partial_x \phi_i(x)}{2\pi} \equiv \sum_i \rho_i(x), \tag{S26}$$

where we used the symmetric basis: $t_i = 1$, and also defined individual channel charge densities $\rho_i = \partial_x \phi_i / (2\pi)$. We can also find the charge density in the diagonal basis from (suppressing the $x$-dependence)

$$\rho = \sum_i \rho_i = \sum_i t_i \frac{\partial_x \phi_i}{2\pi} = \frac{1}{2\pi} \sum_{ijkl} \tilde{t}_i M^{-1}_{ij} M_{jk} \partial_x \tilde{\phi}_k = \frac{1}{2\pi} \sum_{ij} \tilde{t}_i \partial_x \tilde{\phi}_i \equiv \sum_i \tilde{\rho}_i. \tag{S27}$$

The bare mode charge densities can thus be related to the eigenmode charge densities as

$$\rho_i = \sum_{jk} M_{ij} \tilde{T}^{-1}_{jk} \tilde{\rho}_k, \tag{S28}$$

where we defined the diagonal matrix $\tilde{T}_{ij} = \tilde{t}_i \delta_{ij}$. Coupling eigenmodes to local electro-chemical potentials $V_i$, the electrical currents they carry read $I_i = e^2 \tilde{t}_i^2 V_i / h$, so that one may identify "eigenmode conductances" $\tilde{t}_i^2$.

26

In the remainder of this section, we consider the model for the $\nu = 2/3$ edge from the seminal paper by Kane, Fisher, and Polchinski (KFP) [22]. In this model, there are only two bosons ($N = 2$) and

$$K = \begin{pmatrix} 1 & 0 \\ 0 & -3 \end{pmatrix}, \quad V = \begin{pmatrix} v_1 & u \\ u & 3v_2 \end{pmatrix}. \tag{S29}$$

This model describes two interacting and counter-propagating edge modes with filling factor discontinuities $1$ and $-1/3$ respectively.

To diagonalize this Hamiltonian, we first note that $K$ is already on diagonal form. Hence, we may take $\Lambda_1 = \mathbb{1}$. To scale the eigenvalues, we choose $\Lambda_2 = \mathrm{diag}(1, 1/\sqrt{3})$. The remaining step is to diagonalize $V$ while at the same time preserving $\eta = \mathrm{diag}(1, -1)$. To this end, we make the following ansatz for the orthogonal matrix $\Lambda_3$, parametrized by a single parameter $\theta$:

$$\Lambda_3 = \begin{pmatrix} \cosh\theta & -\sinh\theta \\ -\sinh\theta & \cosh\theta \end{pmatrix}. \tag{S30}$$

We then find

$$M = \begin{pmatrix} \cosh\theta & -\sinh\theta \\ -\sinh(\theta)/\sqrt{3} & \cosh(\theta)/\sqrt{3} \end{pmatrix}. \tag{S31}$$

Demanding that $M^T V M$ is diagonal (recall that $M = \Lambda_1 \Lambda_2 \Lambda_3$), we find that $\theta$ must obey

$$\tanh 2\theta = \frac{2u}{\sqrt{3}} (v_1 + v_2)^{-1} \equiv c, \tag{S32}$$

where $c$ is the single number parametrizing edge interactions. It is useful to rephrase this parameter in terms of

$$\Delta \equiv 2\cosh 2\theta - \sqrt{3}\sinh 2\theta = \frac{2 - \sqrt{3}c}{\sqrt{1 - c^2}}. \tag{S33}$$

Then, $\Delta = 2$ corresponds to vanishing interactions, while the special value $\Delta = 1$ is known as the KFP fixed point. The electrical conductances carried by the two eigenmodes are

$$g_\pm \equiv \tilde{t}^2_{i=\pm} = (\Delta \pm 1)/3. \tag{S34}$$

In the absence of interactions, the conductances reduce to the filling factor discontinuities $1$ and $1/3$ respectively. At the KFP fixed point the conductances reflect the edge de-coupling into one charged, $g_+ = 2/3$, and one neutral, $g_- = 0$, mode.

**Charge tunneling on the interacting $\nu = 2/3$ edge**

We now compute the charge tunneling current between two interacting edge modes with $\nu_1 = 1$ and $\nu_2 = 1/3$, following the the approach in Refs. [23–25]. Our starting point is the Hamiltonian obtained from combining Eqs. (S17) and (S29):

$$H_0 = \frac{1}{4\pi} \int dx \left[ \frac{v_1}{\nu_1}(\partial_x \phi_1)^2 + \frac{v_2}{\nu_2}(\partial_x \phi_2)^2 + u(\partial_x \phi_1)(\partial_x \phi_2) \right], \tag{S35}$$

The two bosonic fields obey the commutation relations

$$[\phi_{1,2}(x), \phi_{1,2}(x')] = \pm i\pi\nu_{1,2}\mathrm{sgn}(x - x'). \tag{S36}$$

We next introduce chemical potentials which drive the tunneling current, and take the potentials conjugate to the bare charge densities for these modes. The appropriate term to add to $H_0$ then reads

$$H_V = -\frac{1}{2\pi} \int dx \, [\mu_1 \partial_x \phi_1 + \mu_2 \partial_x \phi_2] = -\int dx \, [\mu_1 \rho_1 + \mu_2 \rho_2]. \tag{S37}$$

Finally, we add a point tunneling Hamiltonian at $x = 0$ by

$$H_\tau = \int dx \, \delta(x) \left[ \Gamma_0 \psi_1^\dagger \psi_2 + h.c. \right]. \tag{S38}$$

Here, $\Gamma_0$ is the tunneling amplitude, and $\psi_1^\dagger(x) = \exp(-i\phi_1(x)/\nu_1)/\sqrt{2\pi b}$ creates an electron in mode 1 at position $x$. Here, $b$ is our UV spatial cut-off. The corresponding creation operator for mode 2 is $\psi_2^\dagger(x) = \exp(i\phi_2(x)/\nu_2)/\sqrt{2\pi b}$, which differs in sign in the exponential from $\psi_1^\dagger$ due to the commutation relations (S36).

Since $H_0$ involves interactions, we diagonalize with $M$ in Eq. (S31). In the diagonal basis, we obtain

$$H_0 = \frac{1}{4\pi} \int dx \left[ v_+(\partial_x \phi_+)^2 + v_-(\partial_x \phi_-)^2 \right], \tag{S39a}$$

$$H_V = -\int dx \, [\mu_+ \rho_+ + \mu_- \rho_-], \tag{S39b}$$

$$H_\tau = \frac{1}{2\pi b} \int dx \, \delta(x) \left[ \Gamma_0 e^{-i\frac{\phi_1}{\nu_1} - i\frac{\phi_2}{\nu_2}} + h.c. \right] = \frac{1}{2\pi b} \int dx \, \delta(x) \left[ \Gamma_0 e^{i\tilde{\boldsymbol{m}}^T \cdot \tilde{\boldsymbol{\phi}}} + h.c. \right], \tag{S39c}$$

where in $H_\tau$, $\tilde{\boldsymbol{m}}$ is obtained from Eq. (S25e) using $\boldsymbol{m} = (-1, -3)$. Moreover, $\tilde{\boldsymbol{\phi}} \equiv (\phi_+, \phi_-)$. We have also re-written $H_V$ in terms of new chemical potentials coupling to the eigen-densities. The bare and eigenmode potentials are related by the linear transformation $(\mu_+, \mu_-) = (\tilde{T}^T)^{-1} M^T (\mu_1, \mu_2)$.

The eigenmodes obey

$$[\phi_{+,-}(x), \phi_{+,-}(x')] = \pm i\pi\mathrm{sgn}(x - x'), \tag{S40}$$

28

and it follows that
$$[\rho_{+,-}(x), \phi_{+,-}(x')] = \pm it_{\pm}\delta(x - x'), \tag{S41}$$

where $\rho_{\pm} = t_{\pm}\partial_x\phi_{\pm}/(2\pi)$, and $t_{\pm}$ are the components of the charge vector in the diagonal basis. They are given by Eq. (S34). The eigenmode tunneling current operator is defined as

$$I_\tau^{\text{em}} \equiv e\frac{d}{dt}\left(\int dx\rho_+\right) = \frac{e}{i\hbar}\left[\int dx\rho_+, H_0 + H_V + H_\tau\right] = \frac{e}{i\hbar}\left[\int dx\rho_+, H_\tau\right], \tag{S42}$$

where we used that $\rho_+$ commutes with the quadratic part of the Hamiltonian, $H_0 + H_V$. Performing the last commutator, we obtain

$$I_\tau^{\text{em}} = \frac{e}{i\hbar}\sqrt{\frac{\Delta^2 - 1}{3}}\sum_{\epsilon=\pm}\epsilon\left[\Gamma_0 e^{i\tilde{\boldsymbol{m}}^T\cdot\tilde{\boldsymbol{\phi}}}\right]_{x=0}^{\epsilon}, \tag{S43}$$

where we have defined

$$\left[\Gamma_0 e^{i\tilde{\boldsymbol{m}}^T\cdot\tilde{\boldsymbol{\phi}}}\right]^+ = \Gamma_0 e^{i\tilde{\boldsymbol{m}}^T\cdot\tilde{\boldsymbol{\phi}}}, \tag{S44a}$$

$$\left[\Gamma_0 e^{i\tilde{\boldsymbol{m}}^T\cdot\tilde{\boldsymbol{\phi}}}\right]^- = \Gamma_0^* e^{-i\tilde{\boldsymbol{m}}^T\cdot\tilde{\boldsymbol{\phi}}}. \tag{S44b}$$

Next, we go to the interaction picture and treat $H_\tau$ as the interaction Hamiltonian. Then, any operator $A$ evolve in time according to

$$A(t) = e^{i(H_0 + H_V)t/\hbar}Ae^{-i(H_0 + H_V)t/\hbar}. \tag{S45}$$

We shall use this equation with $A = I_\tau^{\text{em}}$ and $A = H_\tau$. We begin by computing the commutators

$$\left[H_V, e^{i\tilde{\boldsymbol{m}}^T\cdot\tilde{\boldsymbol{\phi}}}\right] = \sqrt{\frac{\Delta^2 - 1}{3}}e(V_+ - V_-), \tag{S46}$$

where $eV_+ \equiv \mu_{\pm}$. We also have $[H_V, H_0] = 0$. Armed with these commutators, we use the Campbell-Baker-Hausdorff identity

$$e^X Y e^{-X} = Y e^a, \quad \text{if} \quad [X, Y] = aY, \tag{S47}$$

valid if $a$ is a $c$-number, to obtain

$$H_\tau(t) = \sum_{\epsilon=\pm} e^{-i\epsilon\alpha(V_+ - V_-)t/\hbar}\left[\Gamma_0 e^{i\tilde{\boldsymbol{m}}^T\cdot\tilde{\boldsymbol{\phi}}}\right]_{x=0}^{\epsilon} \tag{S48a}$$

$$I_\tau^{\text{em}}(t) = \frac{e\alpha}{i\hbar}\sum_{\epsilon=\pm}\epsilon e^{-i\epsilon\alpha(V_+ - V_-)t/\hbar}\left[\Gamma_0 e^{i\tilde{\boldsymbol{m}}^T\cdot\tilde{\boldsymbol{\phi}}}\right]_{x=0}^{\epsilon}, \tag{S48b}$$

For convenience, we have defined $\alpha \equiv \sqrt{\frac{\Delta^2-1}{3}}$. Next, we compute the expectation value $\langle I_\tau^{\text{em}}(t)\rangle$. To this end, we use the Keldysh formalism and write

$$\langle I_\tau^{\pm}(t)\rangle = \frac{1}{2}\sum_{\eta=\pm}\left\langle T_K\left[I_\tau^{\pm}(t^\eta)\exp\left(-\frac{i}{\hbar}\int_K dt_1 H_\tau(t_1)\right)\right]\right\rangle. \tag{S49}$$

29

In this expression, the Keldysh time-ordering operator, $T_K$ orders along the Keldysh contour $K$: $-\infty \to +\infty \to -\infty$. We denote the "upper" branch $-\infty \to +\infty$ by $\eta = +$ and lower branch $+\infty \to -\infty$ with $\eta = -$. The ordering acts according to $t_1^- > t_2^+$ for all $t_1$ and $t_2$; $t_1^+ > t_2^+$ for $t_1 > t_2$; and $t_1^- > t_2^-$ for $t_1 < t_2$. Since our $H_\tau$ only depends on a single time argument, we have used a symmetric combination on both branches. Hence the factor of $1/2$. To second order in $\Gamma_0$, we find after some manipulations

$$\langle I_\tau^{\text{em}}(t) \rangle = -\frac{ie\alpha|\Gamma_0|^2}{\hbar^2 b^2} \sum_\eta \eta \int_{-\infty}^{\infty} dt_1 \sin(\omega_0(t-t_1)) e^{G_+^{\eta,-\eta}(t-t_1)/\tilde{m}_+^2} e^{G_-^{\eta,-\eta}(t-t_1)/\tilde{m}_-^2}, \tag{S50}$$

where $\tilde{\boldsymbol{m}} = (\tilde{m}_+, \tilde{m}_-)$, $\omega_0 = \alpha e(V_+ - V_-)/\hbar$, and $G_\pm^{\eta,\eta'}$ are Keldysh ordered Green's functions for the eigenmode branches. We are interested in the eigenmode tunneling conductance at finite temperature $k_B T \gg eV$, and we use the finite-temperature Green's functions

$$G_\pm^{\eta,-\eta}(t) = -\log\left[\frac{\sin(\pi(b + iv_\pm t)/(\hbar v_\pm \beta_\pm))}{\pi b/(\hbar v_\pm \beta_\pm)}\right], \tag{S51}$$

where $\beta_\pm = (k_B T_\pm)^{-1}$ reflect the local temperatures of the eigenmodes. We have used a regularization such that $G_\pm^{\eta,-\eta}(0) = 0$. Combining Eqs. (S50) and (S51), we obtain

$$\langle I_\tau^{\text{em}}(t) \rangle = -\frac{ie\alpha|\Gamma_0|^2}{\hbar^2 b^2} \sum_\eta \eta \int_{-\infty}^{\infty} dt_1 \sin(\omega_0(t-t_1)) \Big(\frac{\pi b/(\hbar v_+ \beta_+)}{\sin(\pi(b + iv_+ t/(\hbar v_+ \beta_+)))}\Big)^{m_+^2} \times \tag{S52}$$

$$\Big(\frac{\pi b/(\hbar v_- \beta_-)}{\sin(\pi(b + iv_- t/(\hbar v_- \beta_-)))}\Big)^{m_-^2}. \tag{S53}$$

Next, we perform the variable change $t_1 \to t_1 - ib\eta/v_- + i\eta\hbar\beta_-/2$, and the integral becomes

$$\langle I_\tau^{\text{em}}(t) \rangle = -\frac{ie\alpha|\Gamma_0|^2}{\hbar^2 b^2} \sum_{\eta=\pm 1} \eta \int_{-\infty+ib\eta/v_- - i\eta\hbar\beta_-/2}^{\infty+ib\eta/v_- - i\eta\hbar\beta_-/2} dt_1 \sin\left[\omega_0(t_1 + i\eta\hbar\beta_-/2)\right]$$

$$\times \left(\frac{\pi b/\hbar v_+ \beta_+}{\sin\left[\frac{\pi}{\hbar v_+ \beta_+}\{b(1 - \frac{v_+}{v_-}) - i\eta v_+ t_1 + \frac{\hbar\beta_- v_+}{2}\}\right]}\right)^{m_+^2} \left(\frac{\pi b/\hbar v_- \beta_-}{\sin\left(\frac{\pi}{2} - \frac{i\eta\pi t_1}{\hbar\beta_-}\right)}\right)^{m_-^2}$$

$$= -\frac{ie\alpha|\Gamma_0|^2}{\hbar^2 b^2} \sum_{\eta=\pm 1} \eta \int_{-\infty}^{\infty} dt_1 \sin\left[\omega_0(t_1 + i\eta\hbar\beta_-/2)\right] \left(\frac{\pi b/\hbar v_- \beta_-}{\cosh(\pi t_1/\hbar\beta_-)}\right)^{m_-^2} \left(\frac{\pi b/\hbar v_+ \beta_+}{\sin\left(\frac{\pi\beta_-}{2\beta_+} - \frac{i\pi\eta t_1}{\hbar\beta_+}\right)}\right)^{m_+^2}. \tag{S54}$$

In the second equality of Eq. (S54), the integral limits are changed back to $-\infty$ to $\infty$ because in the integrand of the first right hand side term of Eq. (S54), there is no pole inside the rectangular contour with vertices $t_1 = \infty, -\infty, -\infty + ib\eta/v_- - i\eta\hbar\beta_-/2$, and $\infty + ib\eta/v_- - i\eta\hbar\beta_-/2$ under the condition $\hbar v_- \beta_- > b$. We next target the regime where $|T_+ - T_-| \ll T \equiv (T_+ + T_-)/2$, and obtain the leading term of Eq. (S54)

in $|T_+ - T_-|$ by replacing $T_+$ and $T_-$ by $T$, as

$$\langle I_\tau^{\text{em}}(t)\rangle \simeq \frac{2e\alpha|\Gamma_0|^2}{h^2 b^2}\sinh\left(\frac{\hbar\omega_0}{2k_BT}\right)\int_{-\infty}^\infty dt_1 \cos\left(\omega_0 t_1\right)\left(\frac{\pi bk_BT/\hbar v_-}{\cosh(k_BT\pi t_1/\hbar)}\right)^{m_-^2}\left(\frac{\pi bk_BT/\hbar v_+}{\cosh(k_BT\pi t_1/\hbar)}\right)^{m_+^2}$$

$$= \frac{2e\alpha|\Gamma_0|^2}{h^2 b^2}\left(\frac{\pi bk_BT}{\hbar v_-}\right)^{m_-^2}\left(\frac{\pi bk_BT}{\hbar v_+}\right)^{m_+^2}\sinh\left(\frac{\hbar\omega_0}{2k_BT}\right)\int_{-\infty}^\infty dt_1 \frac{\cos[\omega_0 t_1]}{[\cosh(k_BT\pi t_1/\hbar)]^{m_+^2+m_-^2}}$$

$$= \frac{2e\alpha|\Gamma_0|^2}{h^2 b^2}\left(\frac{\pi bk_BT}{\hbar v_-}\right)^{m_-^2}\left(\frac{\pi bk_BT}{\hbar v_+}\right)^{m_+^2}\sinh\left(\frac{\hbar\omega_0}{2k_BT}\right)\frac{|\Gamma(\Delta + i\frac{\hbar\omega_0}{2\pi k_BT})|^2}{\Gamma(2\Delta)}. \quad \text{(S55)}$$

Here, $\Gamma(x)$ is the gamma function and we used the integral equality $\int_{-\infty}^\infty dt \cosh(2yt)/\cosh^{2x} t = 2^{2x-1}\Gamma(x+y)\Gamma(x-y)/\Gamma(2x)$, valid for $\text{Re}\,x > |\text{Re}\,y|$ and $\text{Re}\,x > 0$. Assuming further $\hbar\omega_0 \ll k_BT$, the tunneling current becomes proportional to the voltage difference $(V_+ - V_-)$ as

$$\langle I_\tau^\pm(t)\rangle = g_{\text{em}}\frac{e^2}{h}(V_+ - V_-), \quad \text{(S56)}$$

with

$$g_{\text{em}} = \frac{\Delta^2 - 1}{3}\frac{|\Gamma_0|^2}{b^2}\frac{|\Gamma(\Delta)|^2}{\Gamma(2\Delta)}\frac{(2\pi)^{2\Delta-2}}{(k_BT)^2}\left(\frac{bk_BT}{\hbar v_-}\right)^{\Delta+1}\left(\frac{bk_BT}{\hbar v_+}\right)^{\Delta-1}. \quad \text{(S57)}$$

This tunneling conductance scales as $T^{2\Delta-2}$ for $\hbar\omega_0 \ll k_BT$ and vanishes as $\Delta - 1$ when $\Delta \to 1$ (cf. Ref. [26]).

**Heat tunneling on the interacting $\nu = 2/3$ edge**

Next, we compute tunneling of energy between the eigenmodes. The energy current operator for tunneling between the eigenmodes is defined as

$$J_\tau^{\text{em}} = -\frac{1}{2}\frac{d}{dt}(H_+ - H_-) = \frac{i}{2\hbar}\left[H_+ - H_-, H_+ + H_- + H_\tau\right] = \frac{i}{2\hbar}\left[H_+ - H_-, H_\tau\right], \quad \text{(S58)}$$

where

$$H_+ = \frac{\hbar v_+}{4\pi}\int(\partial_x\phi_+)^2 dx - \int \mu_+\rho_+ dx, \quad H_- = \frac{\hbar v_-}{4\pi}\int(\partial_x\phi_-)^2 dx - \int \mu_-\rho_- dx. \quad \text{(S59)}$$

Performing the commutators, we find

$$J_\tau^\pm = \frac{i}{4}\sum_{\epsilon=\pm}\epsilon\{m_+v_+\partial_x\phi_+(0) + m_-v_-\partial_x\phi_-(0) - \alpha\frac{\mu_+ + \mu_-}{\hbar}, [\Gamma_0\Psi_1^\dagger\Psi_2]^\epsilon\}, \quad \text{(S60)}$$

where $\{\,,\}$ is the anti-commutator. The tunneling energy current operator $J_{\tau,H_0+H_V}^{\text{em}}$ in the interaction picture with respect to $H_0 + H_V$ is related to operators in the interaction picture with respect to $H_0 = H_+ + H_-$ (denoted as subscript $H_0$) as

$$J_{\tau,H_0+H_V}^{\text{em}} = \frac{i}{4}\sum_{\epsilon=\pm}\epsilon e^{-i\epsilon\omega_0 t}\{m_+v_+\partial_x\phi_+(0) + m_-v_-\partial_x\phi_-(0) - \alpha\frac{\mu_+ + \mu_-}{\hbar}, [\Gamma_0\Psi_{1,H_0}^\dagger\Psi_{2,H_0}]^\epsilon\}. \quad \text{(S61)}$$

31

In the Keldysh formalism, the expectation value of the tunneling energy current reads

$$\langle J_\tau^{\text{em}}(t)\rangle = \frac{1}{2}\sum_{\eta=\pm 1}\langle T_C J_{\tau,H_0+H_V}^{\pm}(t^\eta)e^{-\frac{i}{\hbar}\int_C dt_1 H_{\tau,H_0+H_V}(t_1^{\eta_1})}\rangle. \tag{S62}$$

To second order in $\Gamma_0$ we have

$$\langle J_\tau^{\text{em}}(t)\rangle \simeq \frac{|\Gamma_0|^2}{8\hbar}\sum_{\eta,\eta_1=\pm 1}\sum_{\epsilon=\pm}\epsilon\eta_1\int_{-\infty}^{\infty}dt_1 e^{-i\epsilon\omega_0(t-t_1)}\langle T_C\{m_+v_+\partial_x\phi_+(0)+m_-v_-\partial_x\phi_-(0)$$
$$-\alpha\frac{\mu_++\mu_-}{\hbar},[\Psi_{1,H_0}^\dagger(t^\eta)\Psi_{2,H_0}(t^\eta)]^\epsilon\}[\Psi_{1,H_0}^\dagger(t_1^{\eta_1})\Psi_{2,H_0}(t_1^{\eta_1})]^{-\epsilon}\rangle, \tag{S63}$$

and the energy current can be decomposed into $\langle J_\tau^{\text{em,Q}}\rangle$ and $\langle J_\tau^{\text{em,N}}\rangle$ as

$$\langle J_\tau^{\text{em}}(t)\rangle = \langle J_\tau^{\text{em,Q}}(t)\rangle + \langle J_\tau^{\text{em,N}}(t)\rangle, \tag{S64}$$

where

$$\langle J_\tau^{\text{em,Q}}(t)\rangle = -\frac{|\Gamma_0|^2}{8\hbar}\sum_{\eta,\eta_1=\pm 1}\sum_{\epsilon=\pm}\epsilon\eta_1\int_{-\infty}^{\infty}dt_1 e^{i\epsilon\omega_0(t-t_1)}\langle T_C\{m_+\partial_x\phi_+(0)+m_-\partial_x\phi_-(0), \tag{S65}$$
$$[\Psi_{1,H_0}^\dagger(t^\eta)\Psi_{2,H_0}(t^\eta)]^\epsilon\}\times[\Psi_{1,H_0}^\dagger(t_1^{\eta_1})\Psi_{2,H_0}(t_1^{\eta_1})]^{-\epsilon}\rangle,$$

$$\langle J_\tau^{\text{em,N}}(t)\rangle = \alpha\frac{|\Gamma_0|^2(\mu_++\mu_-)}{4\hbar^2}\sum_{\eta,\eta_1=\pm 1}\sum_{\epsilon=\pm}\epsilon\eta_1\int_{-\infty}^{\infty}dt_1 e^{i\epsilon\omega_0(t-t_1)}$$
$$\times \langle T_C[\Psi_{1,H_0}^\dagger(t^\eta)\Psi_{2,H_0}(t^\eta)]^\epsilon[\Psi_{1,H_0}^\dagger(t_1^{\eta_1})\Psi_{2,H_0}(t_1^{\eta_1})]^{-\epsilon}\rangle. \tag{S66}$$

The contribution $\langle J_\tau^{\text{em,Q}}\rangle$ corresponds to the pure tunneling heat current, i.e., for $e(V_+ - V_-) = 0$, while $\langle J_\tau^{\text{em,N}}\rangle$ corresponds to the energy flow in the process of transferring an electron. Proceeding as for the charge current, we obtain $\langle J_\tau^{\text{em,N}}(t)\rangle$ as

$$\langle J_\tau^{\text{em,N}}(t)\rangle = \alpha\frac{i(\mu_++\mu_-)|\Gamma_0|^2}{2h^2b^2}\sum_{\eta,\eta_1=\pm 1}\eta_1\int_{-\infty}^{\infty}dt_1 \sin\left(\omega_0(t-t_1)\right)e^{G_+^{\eta\eta_1}(t-t_1)/m_+^2}e^{G_-^{\eta\eta_1}(t-t_1)/m_-^2}$$
$$\simeq \frac{e}{2h}g_{\text{em}}(\mu_++\mu_-)(V_+ - V_-) = g_{\text{em}}\frac{e^2}{2h}\left(V_+^2 - V_-^2\right). \tag{S67}$$

Using the relation $\langle B_1 e^{B_2}\rangle = \langle B_1 B_2\rangle e^{\langle B_2^2\rangle/2}$ where $B_1$ and $B_2$ are linear operators of free bosons, $\langle J_\tau^{\text{em,Q}}(t)\rangle$ can be written in terms of Green's functions of the modes as

$$\langle J_\tau^{\text{em,Q}}(t)\rangle = -\frac{i|\Gamma_0|^2}{4\pi h b^2}\sum_{\eta,\eta_1=\pm 1}\eta_1\int_{-\infty}^{\infty}dt_1 \cos\left(\omega_0(t-t_1)\right)e^{G_-^{\eta\eta_1}(t-t_1)/m_-^2}e^{G_+^{\eta\eta_1}(t-t_1)/m_+^2}$$
$$\times \left(\frac{v_+}{m_+^2}\partial_x G_+^{\eta\eta_1}(t-t_1)(x,t-t_1)|_{x=0} + \frac{v_-}{m_-^2}\partial_x G_-^{\eta\eta_1}(x,t-t_1)|_{x=0}\right)$$
$$= -\frac{\pi|\Gamma_0|^2}{2h^2b^2}\sum_{\eta=\pm 1}\int_{-\infty}^{\infty}dt'\cos\left(\omega_0 t'\right)\left(\frac{\pi b/\hbar v_-\beta_-}{\sin[\frac{\pi}{\hbar v_-\beta_-}(b-i\eta v_-t')]}\right)^{m_-^2}\left(\frac{\pi b/\hbar v_+\beta_+}{\sin[\frac{\pi}{\hbar v_+\beta_+}(b-i\eta v_+t')]}\right)^{m_+^2}$$
$$\times \left(\frac{1}{m_+^2\beta_+}\frac{\cos[\frac{\pi}{\hbar v_+\beta_+}(b-i\eta v_+ t')]}{\sin[\frac{\pi}{\hbar v_+\beta_+}(b-i\eta v_+ t')]} - \frac{1}{m_-^2\beta_-}\frac{\cos[\frac{\pi}{\hbar v_-\beta_-}(b-i\eta v_-t')]}{\sin[\frac{\pi}{\hbar v_-\beta_-}(b-i\eta v_-t')]}\right). \tag{S68}$$

The last equality comes from the fact that only the $\eta \times \eta_1 = -1$ contribution to the integral survives. Next, we make a change of variables $t' \to t' - ib\eta/v_- + i\eta\hbar\beta_-/2$ and the integral becomes simplified as

$$\langle J_\tau^{\text{em},Q}(t)\rangle = -\frac{\pi|\Gamma_0|^2}{2h^2b^2} \sum_{\eta=\pm 1} \int_{-\infty+ib\eta/v_- -i\eta\hbar\beta_-/2}^{\infty+ib\eta/v_- -i\eta\hbar\beta_-/2} dt' \cos\left[\omega_0(t'+i\eta\hbar\beta_-/2)\right] \left(\frac{\pi b/\hbar v_-\beta_-}{\sin\left(\frac{\pi}{2} - \frac{i\eta\pi t'}{\hbar\beta_-}\right)}\right)^{m_-^2}$$

$$\times \left(\frac{\pi b/\hbar v_+\beta_+}{\sin\left[\frac{\pi}{\hbar v_+\beta_+}\left\{b(1-\frac{v_+}{v_-}) - i\eta v_+ t' + \frac{\hbar\beta_-v_+}{2}\right\}\right]}\right)^{m_+^2}$$

$$\times \left(\frac{1}{m_+^2\beta_+}\frac{\cos\left[\frac{\pi}{\hbar v_+\beta_+}\{b(1-\frac{v_+}{v_-}) - i\eta v_+ t' + \frac{\hbar\beta_-v_+}{2}\}\right]}{\sin\left[\frac{\pi}{\hbar v_+\beta_+}\{b(1-\frac{v_+}{v_-}) - i\eta v_+ t' + \frac{\hbar\beta_-v_+}{2}\}\right]} - \frac{1}{m_-^2\beta_-}\frac{\cos\left(\frac{\pi}{2} - \frac{i\eta\pi t'}{\hbar\beta_-}\right)}{\sin\left(\frac{\pi}{2} - \frac{i\eta\pi t'}{\hbar\beta_-}\right)}\right)$$

$$= -\frac{\pi|\Gamma_0|^2}{2h^2b^2}\sum_{\eta=\pm 1}\int_{-\infty}^{\infty} dt' \cos\left[\omega_0(t'+i\eta\hbar\beta_-/2)\right]\left(\frac{\pi b/\hbar v_-\beta_-}{\cosh(\pi t'/\hbar\beta_-)}\right)^{m_-^2}$$

$$\times \left(\frac{\pi b/\hbar v_+\beta_+}{\sin\left(\frac{\pi\beta_-}{2\beta_+} - \frac{i\pi\eta t'}{\hbar\beta_+}\right)}\right)^{m_+^2}\left(\frac{1}{m_+^2\beta_+\tan\left(\frac{\pi\beta_-}{2\beta_+} - \frac{i\pi\eta t'}{\hbar\beta_+}\right)} - \frac{i\eta}{\beta_- m_-^2}\tanh(\frac{\pi t'}{\hbar\beta_-})\right). \tag{S69}$$

We next compute $\langle J_\tau^{\text{em},Q}\rangle$ to first order in $\Delta T = T_+ - T_-$ in the limit of $|T_+ - T_-| \ll T = (T_+ + T_-)/2$. In addition, we ignore the terms of second order in $\omega_0$ (there is no first order due to the cosine). We then obtain

$$\langle J_\tau^{\text{em},Q}\rangle \simeq \frac{\pi^{2+m_+^2+m_-^2}m_-^2|\Gamma_0|^2 k_B \Delta T}{4h^2 b^2}\left(\frac{bk_BT}{\hbar v_+}\right)^{m_+^2}\left(\frac{bk_BT}{\hbar v_-}\right)^{m_-^2}$$

$$\int_{-\infty}^{\infty} dt'\left(2 + m_+^2 - m_-^2 + (m_+^2 - m_-^2)\cosh(\frac{2\pi t k_B T}{\hbar})\right)\cosh(\frac{\pi t k_B T}{\hbar})^{-2-m_+^2-m_-^2}$$

$$= \frac{\hbar m_-^2 m_+^2 \pi^{3/2+m_-^2+m_+^2}|\Gamma_0|^2 k_B\Delta T\Gamma(m_-^2/2 + m_+^2/2)}{2bh^2 k_B T\Gamma(\frac{1}{2}(3+m_-^2+m_+^2))}\left(\frac{bk_BT}{\hbar v_+}\right)^{m_+^2}\left(\frac{bk_BT}{\hbar v_-}\right)^{m_-^2}$$

$$= g_{\text{em}}\frac{\pi^2 k_B^2(T_+^2 - T_-^2)}{6h}\frac{9}{1+2\Delta}. \tag{S70}$$

Thus, we find for heat tunneling

$$\langle J_\tau^{\text{em}}(t)\rangle = g_{\text{em}}\frac{e^2}{2h}\left(V_+^2 - V_-^2\right) + g_{\text{em}}\frac{\pi^2 k_B^2(T_+^2 - T_-^2)}{6h}\frac{9}{1+2\Delta} \tag{S71}$$

which is the total heat current in the linear response regime. We note that interactions renormalizes the Wiedemann-Franz parameter for heat tunneling [24]

$$\gamma = 9/(1+2\Delta). \tag{S72}$$

The approximations we have made are good when

$$(|\Gamma_0|/b) \ll k_B T \ll (\hbar v_\pm/b), \tag{S73}$$

i.e. when the average temperature $T$ is smaller than the bandwidth $\hbar v/b$ of the modes [24]. Moreover, these scales must be much larger than the tunneling energy. This corresponds to weak tunneling.

33

The key results of the micropscopic calculations are Eqs. (S56), (S57), and (S71). We shall next use them to derive equilibration lengths for charge and heat equilibration.

**Charge transport and equilibration**

The discussion in this subsection parallels that in Ref. [26] (co-authored by two of the present authors), whose primary focus is on the effect of the contact region on charge transport.

To model charge equilibration, we first introduce electronic scattering rates $\Gamma_\pm$ between the eigenmode charge densities. We also recall the two dimensionless conductances for the eigenmodes: $g_\pm = \tilde{t}_\pm^2 = (\Delta \pm 1)/3$. Then, we can re-express $\alpha^2 = (\Delta^2 - 1)/3 = 3g_+ g_-$. Note that in the limit $\Delta \to 1$, the charge carried by $\tilde{\phi}_-$ vanishes.

We assume that that successive scattering events are incoherent: $k_B T > \hbar v_\pm /a$ where $a$ is the typical distance between scatterers. We can then write steady state equations of motion on the $\nu = 2/3$ edge away from contacts (i.e., with vanishing source terms) as

$$\partial_x \begin{pmatrix} v_+ \rho_+ \\ v_- \rho_- \end{pmatrix} = \begin{pmatrix} -\Gamma_+ & \Gamma_- \\ -\Gamma_+ & \Gamma_- \end{pmatrix} \begin{pmatrix} \rho_+ \\ \rho_- \end{pmatrix}. \tag{S74}$$

The first and second rows in the matrix in the right-hand-side of Eq. (S74) are identical in view of current conservation. We use now relations between local currents, densities, and voltages of the eigenmodes:

$$I_\pm = v_\pm \rho_\pm = \frac{e^2}{h} g_\pm V_\pm. \tag{S75}$$

Using the first equality in Eq. (S75), we rewrite Eq. (S74) in terms of currents:

$$\partial_x \begin{pmatrix} I_+ \\ I_- \end{pmatrix} = \begin{pmatrix} -\Gamma_+/v_+ & \Gamma_-/v_- \\ -\Gamma_+/v_+ & \Gamma_-/v_- \end{pmatrix} \begin{pmatrix} I_+ \\ I_- \end{pmatrix}. \tag{S76}$$

In view of gauge invariance, the right-hand-side of Eq. (S74) should depend on potentials $V_\pm$ through their difference only. In combination with the second equality in Eq. (S75), this implies that

$$\frac{\Gamma_+}{\Gamma_-} = \frac{g_- v_+}{g_+ v_-}. \tag{S77}$$

The inter-channel tunneling current per unit length reads

$$\langle I_\tau^{\text{em}} \rangle = \frac{e^2}{h} \frac{g_{\text{em}}}{a} (V_+ - V_-) = \Gamma_+ \rho_+ - \Gamma_- \rho_-. \tag{S78}$$

34

The first form here follows from Eq. (S56), while the second one from Eq. (S74). We thus express $\Gamma_\pm$ through $g_{\rm em}$:

$$\frac{\Gamma_\pm}{v_\pm} = \frac{g_{\rm em}}{a g_\pm}. \tag{S79}$$

Using Eq. (S57) for $g_{\rm em}$ and substituting the resulting $\Gamma_\pm/v_\pm$ into Eq. (S76), we obtain the transport equation for the edge currents in the form

$$\partial_x \begin{pmatrix} I_+ \\ I_- \end{pmatrix} = \frac{1}{\ell_{\rm eq}^C} \begin{pmatrix} -g_- & g_+ \\ -g_- & g_+ \end{pmatrix} \begin{pmatrix} I_+ \\ I_- \end{pmatrix}, \tag{S80}$$

where

$$\ell_{\rm eq}^C = \frac{a g_+ g_-}{g_{\rm em}} = \frac{a}{3} \times \left[ \frac{|\Gamma_0|^2}{b^2} \frac{|\Gamma(\Delta)|^2}{\Gamma(2\Delta)} \frac{(2\pi)^{2\Delta-2}}{(k_B T)^2} \left(\frac{b k_B T}{\hbar v_-}\right)^{\Delta+1} \left(\frac{b k_B T}{\hbar v_+}\right)^{\Delta-1} \right]^{-1} \tag{S81}$$

defines the corresponding equilibration length. By using the relations (S75), we can equivalently rewrite Eq. (S82) as an equation on local voltages:

$$\partial_x \begin{pmatrix} V_+ \\ V_- \end{pmatrix} = \frac{1}{\ell_{\rm eq}^C} \begin{pmatrix} -g_- & g_- \\ -g_+ & g_+ \end{pmatrix} \begin{pmatrix} V_+ \\ V_- \end{pmatrix}. \tag{S82}$$

To solve Eq. (S82), we must supplement it with boundary conditions set by the contacts. The simplest assumption is that interactions are fully screened in the regions covered by contacts. Equivalently, this corresponds to the assumption that in the vicinity of contacts 1 and 1/3 modes are eigenmodes [27]. In that case, in a two-terminal geometry with length $L$, the solution for the charge conductance $G$ reads

$$G = \frac{2}{3} \frac{e^2}{h} \times \frac{3 + e^{-2L/3\ell_{\rm eq}^C}}{3 - e^{-2L/3\ell_{\rm eq}^C}}, \tag{S83}$$

and the only dependence of $G$ on $\Delta$ is in $\ell_{\rm eq}^C$. For $L \ll \ell_{\rm eq}^C$ one finds $G \simeq 4e^2/3h$, while for $L \gg \ell_{\rm eq}^C$ the conductance is given by its equilibrated value, $G \simeq 2e^2/3h$, up to exponentially small corrections in $L/\ell_{\rm eq}^C$. A detailed study of more general contact models leads to qualitatively similar results [26]. The crossover of the charge conductance from 4/3 to 2/3 (in units of $e^2/h$) with increasing system size was predicted in Ref. [27] and observed in Ref. [28] in artificially engineered 2/3 edges.

The exponential approach to the equilibrated value of the conductance at $L/\ell_{\rm eq}^C$ reflects the chiral character of the charge transport ($g_+ \neq g_-$). The rate of the exponential approach is determined by a non-zero eigenvalue of the matrix in Eq. (S82).

It is important to emphasize that, while the prefactor in Eq. (S57) for $g_{\rm em}$ is proportional to $\Delta - 1$ at $\Delta \to 1$, this singularity is canceled in $\ell_{\rm eq}^C$ by $g_-$ which is also proportional to $\Delta - 1$, see Eq. (S34). Therefore, $\ell_{\rm eq}^C$ remains finite in the limit $\Delta \to 1$ (cf. Ref. [26]). Below, we contrast this result with the behavior of the heat equilibration length which diverges in the limit $\Delta \to 1$.

35

**Heat transport and equilibration**

To model the heat transport, we use the same approach as in the previous section. We introduce the energy scattering rates $\Gamma_\pm^H$ between the eigenmodes and heat currents carried by eigenmodes, $J_\pm = v_\pm n_\pm$, where $n_\pm$ are the eigenmode energy densities. These heat currents can be related to effective temperatures as $J_\pm = \frac{1}{2}\kappa_0 T_\pm^2$ (with $\kappa_0 = \pi^2 k_B^2/3h$). For the tunneling heat current between the eigenmodes we then have

$$\langle J_\tau^{\rm em}\rangle = \frac{g_{\rm em}}{a}\gamma\frac{\kappa_0(T_+^2 - T_-^2)}{2} = \Gamma_+^H n_+ - \Gamma_-^H n_-, \tag{S84}$$

which is the counterpart of Eq. (S78). Here we used Eq. (S71) with $V_+ - V_- = 0$ (pure heat transport). The transport equation is now derived in full analogy with the case of charge transport, and we obtain

$$\partial_x \begin{pmatrix} J_+ \\ J_- \end{pmatrix} = \frac{1}{\ell_{\rm eq}^H}\begin{pmatrix} -1 & 1 \\ -1 & 1 \end{pmatrix}\begin{pmatrix} J_+ \\ J_- \end{pmatrix}, \tag{S85}$$

which is the counterpart of Eq. (S82). Equivalently, this equation can be written in terms of local temperatures

$$\partial_x \begin{pmatrix} T_+^2(x) \\ T_-^2(x) \end{pmatrix} = \frac{1}{\ell_{\rm eq}^H}\begin{pmatrix} -1 & 1 \\ -1 & 1 \end{pmatrix}\begin{pmatrix} T_+^2(x) \\ T_-^2(x) \end{pmatrix}. \tag{S86}$$

Here, $\ell_{\rm eq}^H$ is the thermal equilibration length

$$\ell_{\rm eq}^H = \frac{a}{g_{\rm em}\gamma} = \frac{\ell_{\rm eq}^C}{(\Delta^2 - 1)\gamma/3}. \tag{S87}$$

The temperature scaling of both equilibration lengths, $\ell_{\rm eq}^C$ and $\ell_{\rm eq}^H$, is $T^{2-2\Delta}$, in agreement with Ref. [27] (where, however, no distinction between the two lengths was made).

In contrast to Eq. (S82), the matrix in Eq. (S85) has two zero eigenvalues. This reflects the fact that the heat conductances of the two eigenmodes are identical (equal to $\kappa_0 T$) and independent of their electrical charges, cf. Ref. [29]. The transport equation solutions are then linear in $x$, which implies heat diffusion along the edge.

We consider now the two-terminal geometry for heat transport, with contacts "biased" with temperatures $T$ and $T + \Delta T$, respectively. To stay in the linear-transport regime, we assume $\Delta T \ll T$. Solving Eq. (S85), we find the two-terminal heat conductance

$$G^H = 2\kappa_0 T \frac{1}{1 + L/\ell_{\rm eq}^H}, \tag{S88}$$

which yields the ballistic heat conductance $G^H \simeq 2\kappa_0 T$ for $L \ll \ell_{\rm eq}^H$ and diffusive heat conductance $G^H \simeq 2\kappa_0 T \ell_{\rm eq}^H/L$ for $L \gg \ell_{\rm eq}^H$.

According to Eq. (S87), the thermal equilibration length diverges as $(\Delta - 1)^{-1}$ at $\Delta \to 1$. Therefore, for strong interactions such as $\Delta$ is close to unity, we have $\ell_{\text{eq}}^H \gg \ell_{\text{eq}}^C$ and thus a a broad regime of system sizes $L$ satisfying $\ell_{\text{eq}}^C \ll L \ll \ell_{\text{eq}}^H$. In this regime, the charge and heat conductances are, according to Eqs. (S83) and (S88),

$$G \approx \frac{2e^2}{3h} \quad \text{and} \quad G^H \approx 2\kappa_0 T, \tag{S89}$$

explaining the experimental observations of the present work.


\end{document}